\pdfoutput=1
\documentclass[11pt,a4paper]{article}
\usepackage{jheppub}

\input{frontmatter}

\newcommand{\ourintlumi}{{4.7\,\ifb{}}}
\newcommand{\multijettitle}{{
Hunt for new phenomena using 
large jet multiplicities and missing transverse momentum
with ATLAS in \boldmath $\ourintlumi$ of
$\sqrt{s}=7\,\TeV{}$ proton-proton collisions
}}

\newcommand\multijetabstract{{
Results are presented of a search for new particles 
decaying to large numbers of jets 
in association with missing transverse momentum,
using \ourintlumi{} of $pp$ collision data 
at $\sqrt{s} = 7\,\TeV$ collected by the ATLAS experiment 
at the Large Hadron Collider in 2011.
The event selection requires missing transverse momentum, 
no isolated electrons or muons,  
and from $\geq$6 to $\geq$9 jets.
No evidence is found for physics beyond the Standard Model.
The results are interpreted in the context of a \mSUGRA{} supersymmetric model, 
where, for large universal scalar mass $m_0$, gluino masses smaller than $840\,\GeV$ are excluded at the 95\% confidence level,
extending previously published limits.
Within a simplified model containing only a gluino octet and a neutralino, 
gluino masses smaller than $870\GeV$ are similarly excluded 
for neutralino masses below $100\,\GeV$.
}}

\usepackage{preprintcover} 
\PreprintCoverPaperTitle{\multijettitle}
\PreprintIdNumber{CERN-PH-EP-2012-141}  
\PreprintCoverAbstract{\multijetabstract}
\PreprintJournalName{Journal of High Energy Physics}

\begin{document}

\title{\multijettitle}
\author{The ATLAS Collaboration}

\abstract{\multijetabstract}

\collaborationImg{\includegraphics{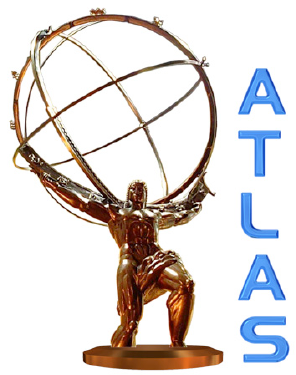}}

\maketitle

\newcommand\hepdatablurb{Acceptance times efficiency values are tabulated elsewhere~\cite{multijet_2012_auxiliary}.}

\section{Introduction} \label{sec:intro}

Many extensions of the Standard Model of particle physics
predict the presence of TeV-scale strongly interacting particles
that decay to lighter, weakly interacting descendants.
Any such weakly interacting particles that are massive and stable can
contribute to the dark matter content of the universe.
The strongly interacting parents would be produced in the proton-proton interactions at the Large Hadron Collider (LHC),
and such events would be characterized by 
significant missing transverse momentum \met{} from the unobserved weakly interacting 
daughters, and jets from emissions of quarks and/or gluons.

In the context of $R$-parity conserving~\cite{Fayet:1976et,Fayet:1977yc,Farrar:1978xj,Fayet:1979sa,Dimopoulos:1981zb}
supersymmetry~\cite{Dimopoulos:1981zb,Witten:1981nf,Dine:1981za,Dimopoulos:1981au,Sakai:1981gr,Kaul:1981hi}, 
the strongly interacting parent particles are the squarks~$\tilde{q}$ and gluinos~$\tilde{g}$,
they are produced in pairs, and the lightest
supersymmetric particles can provide the stable dark matter candidates~\cite{Goldberg:1983nd,Ellis:1983ew}.
Jets are produced from a variety of sources: 
from quark emission in supersymmetric cascade decays,
production of heavy Standard Model particles ($W$, $Z$ or $t$) 
which then decay hadronically, or from QCD radiation.
Examples of particular phenomenological interest include models 
where squarks are significantly heavier than gluinos.
In such models the gluino pair production and decay process 
\begin{equation*}
\tilde{g}+ \tilde{g} \rightarrow \left( t + \bar{t} + \ninoone \right) + \left( t + \bar{t} + \ninoone \right)
\end{equation*}
can dominate, producing large jet multiplicities 
when the resulting top quarks decay hadronically.
In the context of \msugra{} models, 
a variety of different cascade decays, including the $\tilde{g} \tilde{g}$ initiated process above, 
can lead to large jet multiplicities.

A previous ATLAS search in high jet multiplicity final states~\cite{Aad:2011qa} 
examined data taken during the first half of 2011,
corresponding to an integrated luminosity of $1.34\,\ifb$.
This \paper{} extends the analysis
to the complete ATLAS 2011 $pp$ data set, corresponding to \ourintlumi,
and includes improvements in the analysis and event selection 
that further increase sensitivity to models of interest.

Events are selected with large jet multiplicities
ranging from $\geq\,$6 to $\geq\,$9 jets, 
in association with significant \met{}. 
Events containing high transverse momentum ($\pT$) electrons 
or muons are vetoed in order to reduce backgrounds 
from (semi-leptonically) decaying top quarks or $W$ bosons. 
Other complementary searches have been performed by the ATLAS collaboration 
in final states with \met{} and one or more leptons~\cite{ATLAS:2011ad,Aad:2011cwa}.
Further searches have been carried out by ATLAS using events
with at least two, three or four jets~\cite{Aad:2011ib},
or with at least two $b$-tagged jets~\cite{Aad:2011cw}.
Searches have also been performed by the CMS collaboration, including a recent 
analysis in fully hadronic final states~\cite{Chatrchyan:2011zy}.

\section{The ATLAS detector and data samples}\label{sec:samples}

The ATLAS experiment~\cite{Aad:2008zzm} is a multi-purpose particle
physics detector with a forward-backward symmetric cylindrical
geometry and nearly 4$\pi$ coverage in solid angle.\footnote{ATLAS 
uses a right-handed coordinate system with its origin at the nominal
interaction point in the centre of the detector and the $z$-axis along the
beam pipe. Cylindrical coordinates $(r,\phi)$ are used in the transverse
plane, $\phi$ being the azimuthal angle around the beam pipe. The pseudorapidity $\eta$ is
defined in terms of the polar angle $\theta$ by $\eta=-\ln\tan(\theta/2)$.
} The layout of the detector is
dominated by four superconducting magnet systems, which comprise a
thin solenoid surrounding inner tracking detectors,  
and a barrel and two end-cap toroids supporting a large muon spectrometer.
The calorimeters are of particular importance to this analysis.  In the pseudorapidity
region $\left|\eta\right| < 3.2$, high-granularity liquid-argon (LAr)
electromagnetic (EM) sampling calorimeters are used.  An iron/scintillator-tile
calorimeter provides hadronic coverage for 
$\left|\eta\right| < 1.7$.  The end-cap and forward regions, spanning
$1.5 < \left|\eta\right| < 4.9$, are instrumented with LAr calorimetry
for both EM and hadronic measurements.

The data sample used in this analysis was taken during 
April -- October 2011
with the LHC operating at a proton-proton centre-of-mass energy of $\sqrt{s}=7~\TeV$.
Application of beam, detector and data-quality requirements resulted in
a corresponding integrated luminosity of $4.7 \pm 0.2\,\ifb$~\cite{lumi_summer_11}.
The analysis makes use of dedicated multi-jet triggers
that required
either at least four jets with $\pt>45\,\GeV{}$ or at least five jets with $\pt>30\,\GeV{}$, 
where the energy is measured at the electromagnetic scale\footnote{The electromagnetic scale is the basic calorimeter signal scale for the ATLAS calorimeters. It has been established using test-beam measurements for electrons and muons to give the correct response for the energy deposited in electromagnetic showers, although it does not correct for the lower response of the calorimeter to hadrons.}
and the jets must have $|\eta|<3.2$.
In all cases the trigger efficiency was greater than 98\%
for events satisfying the offline jet multiplicity selections
described in \oursecref{sec:eventselection}.

\section{Object reconstruction}\label{sec:objectselection}

The jet, lepton and missing transverse momentum definitions are 
based closely on those of Ref.~\cite{Aad:2011qa},
with small updates to account for evolving accelerator and detector conditions.

Jet candidates are reconstructed using the
anti-$k_t$ jet clustering algorithm~\cite{Cacciari:2008gp,Cacciari:2005hq} with
radius parameter of $0.4$. The inputs to this algorithm are clusters
of calorimeter cells seeded by cells with energy significantly above
the noise level. Jet momenta are reconstructed by performing a
four-vector sum over these topological clusters of calorimeter cells, 
treating each as an
$(E,\vec{p})$ four-vector with zero mass. The jet energies are corrected for
the effects of calorimeter non-compensation and inhomogeneities by
using $\ourpt$- and $\eta$-dependent calibration factors
based on Monte Carlo (MC) simulations validated with extensive
test-beam and collision-data studies~\cite{Aad:2011he}.
Only jet candidates with $\ourpt > 20$
\GeV{} and $|\eta| < 4.9$ are retained.
Further corrections are applied to any jet falling in problematic areas of the calorimeter. 
The event is rejected if, for any jet, this additional correction leads to a contribution to $\met$ that is 
greater than both $10\,\GeV$ and $0.1\,\met$. These criteria, along
with selections against non-collision background and calorimeter
noise, lead to a loss of signal efficiency of $\sim$8\% for the models considered.
When identification of jets containing heavy flavour quarks is required, 
either to make measurements in control regions or for cross checks, 
a tagging algorithm exploiting both impact parameter and secondary vertex information is used.
Jets are tagged for $|\eta| < 2.5$ and the parameters of the algorithm
are chosen such that 70\% of $b$-jets 
and $\sim$1\% of light flavour or gluon jets, are selected in
\ttbar{} events in Monte Carlo simulation~\cite{ATLAS-CONF-2011-102}.
Jets initiated by charm quarks are tagged with about 20\% efficiency.

Electron candidates are required to have $\ourpt$ $>$ $20~\GeV$ and $|\eta|$ $<$
2.47, and to satisfy the `medium' electron shower shape and track selection criteria of
Ref.~\cite{ATLAS:2011ad}.
Muon candidates are required to have $\ourpt>10\GeV$ and $|\eta|$ $<$ 2.4.
Additional requirements are applied to 
muons when defining leptonic control regions. In this case  
muons must have longitudinal and transverse impact parameters within 1\,mm and 0.2\,mm of the primary vertex, respectively, and the sum of the transverse momenta of other tracks within a cone of $\Delta R=0.2$ around the muon must be less than $1.8\,\GeV$,
where $\Delta R = \sqrt{(\Delta \eta)^2 + (\Delta \phi)^2}$.

The measurement of the missing transverse momentum two-vector
$\ourvecptmiss$ and its magnitude (conventionally denoted $\met$) is then based on
the transverse momenta of all electron and muon candidates, all jets 
with $|\eta|<4.5$ which are not also electron candidates, and all calorimeter clusters with $|\eta|<4.5$ not associated to such objects~\cite{Aad:2012re}.

Following the steps above, overlaps between candidate jets with $|\eta|<2.8$ 
and leptons are resolved as follows.
First, any such jet candidate lying within a distance 
$\Delta R =0.2$ of an electron is discarded,
then any lepton candidate remaining within a distance
$\Delta R =0.4$ of such a jet candidate is discarded.
Thereafter, all jet candidates with $|\eta|>2.8$ are discarded, 
and the remaining electron, muon and jet candidates are
retained as reconstructed objects.

\section{Event selection}\label{sec:eventselection}

Following the object reconstruction described in \oursecref{sec:objectselection}, 
events are discarded 
if they contain any jet
failing quality criteria designed to suppress detector noise and
non-collision backgrounds, or if they lack a reconstructed primary
vertex with five or more associated tracks.

\begin{table*}
\renewcommand\arraystretch{1.4}
 \begin{center}
   \begin{tabular}{ | l || c | c | c | c | c | c |}
     \hline
{\bf Signal region} & ~{\bf 7j55}~ & ~{\bf 8j55}~ & {\bf 9j55} & ~{\bf 6j80}~ & ~{\bf 7j80}~ & ~{\bf 8j80}~\\ \hline \hline
     Number of isolated leptons $(e,\,\mu)$ & \multicolumn{6}{| c |}{$=0$} \\ \hline
     Jet $p_{\rm T}$ &  \multicolumn{3}{| c |}{$>$\,55\,\GeV{}} &  \multicolumn{3}{| c | }{$>$\,80\,\GeV{}}  \\ \hline
     Jet $|\eta|$ &  \multicolumn{6}{|c|}{$<$\,2.8} \\ \hline
     Number of jets & $\geq$\,7 & $\geq$\,8 & $\geq$\,9 & $\geq$\,6 & $\geq$\,7 & $\geq$\,8 \\ \hline
     $\met/\sqrt{H_{\rm T}}$ & \multicolumn{6}{|c|}{$>\,4\,\rootgev$}   \\ \hline
   \end{tabular}
   \caption{\label{tab:sr}
     Definitions of the six signal regions.
   }
 \end{center}
\end{table*}

For events containing no isolated electrons or muons, 
six non-exclusive signal regions (SRs) are defined as shown in \ourtabref{tab:sr}.
The first three require at least seven, eight or nine jets, respectively, with $\pt{}>55\,\GeV$;
the latter three require at least six, seven or eight jets, respectively, with $\pt{}>80\,\GeV$.
The final selection variable is $\metsig$, the ratio 
of the magnitude of the missing transverse momentum
to the square root of the scalar sum \HT{} of the transverse momenta 
of all jets with $\pT>40\,\GeV$ and $|\eta|<2.8$.
This ratio is closely related to the significance of the missing transverse momentum
relative to the resolution due to stochastic variations 
in the measured jet energies~\cite{Aad:2012re}.
The value of \metsig{} is required to be larger than 4\,\rootgev{}
for all signal regions.

A previous ATLAS analysis of similar final states~\cite{Aad:2011qa}
required jets to be separated by $\Delta R > 0.6$
to ensure that the trigger efficiency was on its plateau.
It has since been demonstrated that the requirement of an
offline jet multiplicity at least one larger 
than that used in the trigger is sufficient to 
achieve a 98\% trigger efficiency.
Investigations on the enlarged data sample, in comparison to the previous incarnation of the strategy used here, allow various improvements to be made; in particular, the requirement on jet-jet separation is modified so as to increase the acceptance for signal models of interest by a factor two to five, 
without introducing any significant trigger inefficiency.

The dominant backgrounds are multi-jet production,
including purely strong interaction processes and
fully hadronic decays of \ttbar;
semi- and fully-leptonic decays of \ttbar{};
and leptonically decaying $W$ or $Z$ bosons produced in association with jets.
Non-fully-hadronic \ttbar{}, and $W$ and $Z$ are collectively
referred to as \leptonic{} backgrounds.
Contributions from gauge boson pair and single top quark production are negligible.
The determination of the multi-jet and \leptonic{}
backgrounds is described in Sections~\ref{sec:backgrounds:qcd}
and \ref{sec:backgrounds:leptonic}, respectively.

\section{Monte Carlo simulations}\label{sec:backgrounds}\label{sec:mc}

Monte Carlo simulations are used as part of the \leptonic{} background determination process, 
and to assess sensitivity to specific SUSY signal models.
The \leptonic{} backgrounds
are generated using \alpgenVER{2.13}~\cite{Mangano:2002ea}
with the PDF set {\tt CTEQ6L1}~\cite{Pumplin:2002vw}.
Fully-leptonic \ttbar{} events are generated with up to five additional partons
in the matrix element,
while semi-leptonic \ttbar{} events are generated with up to three additional partons
in the matrix element.
$W$~+~jets and $Z\rightarrow \nu\bar\nu$~+~jets are generated with up to six
additional partons, and the 
$Z\rightarrow \ell^+\ell^-$~+~jets (for $\ell\in\{e,\,\mu,\,\tau\}$) process
is generated with up to five additional partons in the matrix element.
In all cases, additional jets are generated via parton showering, which, together with
fragmentation and hadronization, is
performed by \herwig~\cite{Corcella:2000bw,Corcella:2002jc}.
\jimmy~\cite{Butterworth:1996zw} is used to simulate the underlying event.
The $W$~+~jets, $Z$~+~jets and \ttbar{} backgrounds are normalized according to their inclusive theoretical cross sections~\cite{Aliev:2010zk,Melnikov:2006kv}. 
The estimation of the \leptonic{} backgrounds in the signal regions is described in detail in \oursecref{sec:backgrounds:leptonic}.

Supersymmetric production processes are generated using 
\herwigppVER{2.4.2}~\cite{Bahr:2008pv}.
Signal cross sections are calculated to next-to-leading order in the strong coupling constant $\alpha_S$, including the resummation of soft gluon emission at next-to-leading-logarithmic accuracy (NLO+NLL)~\cite{Beenakker:1996ch,Kulesza:2008jb,Kulesza:2009kq,Beenakker:2009ha,Beenakker:2011fu}.\footnote{The NLL correction is used for squark and gluino production when the average of the squark masses in the first two generations and the gluino mass lie between 200~GeV and 2~TeV. In the case of gluino-pair (associated squark-gluino) production processes, the calculations were extended up to squark masses of 4.5 TeV (3.5 TeV). For masses outside this range and for other types of production processes (i.e. electroweak and associated strong and electroweak),
cross sections at NLO accuracy obtained with 
\prospinoVER{2.1}~\cite{Beenakker:1996ch} are used.}
An envelope of cross-section predictions is defined using the 68\% confidence-level (CL) ranges of the {\tt CTEQ6.6}~\cite{Nadolsky:2008zw} 
(including the $\alpha_S$ uncertainty) and 
\mstwVER{2008 NLO}~\cite{Martin:2009iq}
PDF sets, together with independent variations of the factorization and renormalization scales 
by factors of two or one half. 
The nominal cross-section value is then taken to be the midpoint of the envelope, and the uncertainty assigned is half the full width of the envelope, following closely the PDF4LHC recommendations~\cite{Botje:2011sn}.
\msugra{} particle spectra and decay modes are calculated with \isajetVER{7.75}~\cite{Paige:2003mg}.
For illustrative purposes, plots of kinematic quantities 
show the distribution expected for an example \mSUGRA{}
point that has not been excluded in previous searches. 
This reference point is defined
by\footnote{A particular \mSUGRA{} model point is specified by five parameters:
the universal scalar mass $m_0$,
the universal gaugino mass $m_{1/2}$,
the universal trilinear scalar coupling $A_0$,
the ratio of the vacuum expectation values of the two Higgs fields $\tan\beta$,
and the sign of the higgsino mass parameter $\mu$.}:
$m_0=2960\GeV$, $m_{1/2} = 240\GeV$, $A_0=0$, $\tan\beta=10$, and $\mu>0$.

All Monte Carlo samples employ a 
detector simulation~\cite{Aad:2010ah} based on 
{\tt GEANT4}~\cite{Agostinelli:2002hh}
and are reconstructed with the same algorithms as the data.

\section{Multi-jet backgrounds}\label{sec:backgrounds:qcd}

\begin{figure*}
\begin{center}
\subfigure[\label{fig:metsig_vr:55}]{\includegraphics[width=0.45\linewidth]{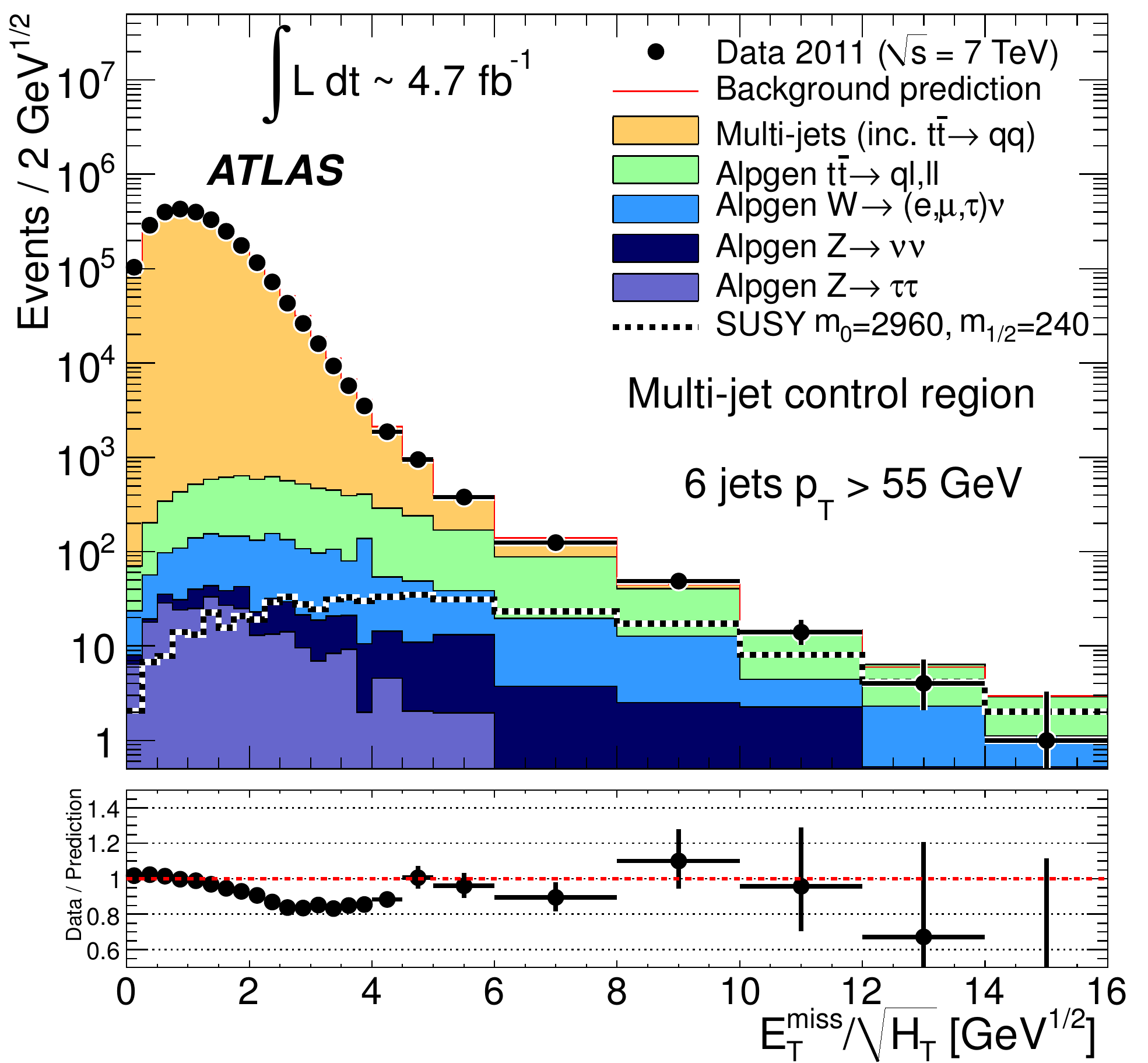} }
\subfigure[\label{fig:metsig_vr:80}]{\includegraphics[width=0.45\linewidth]{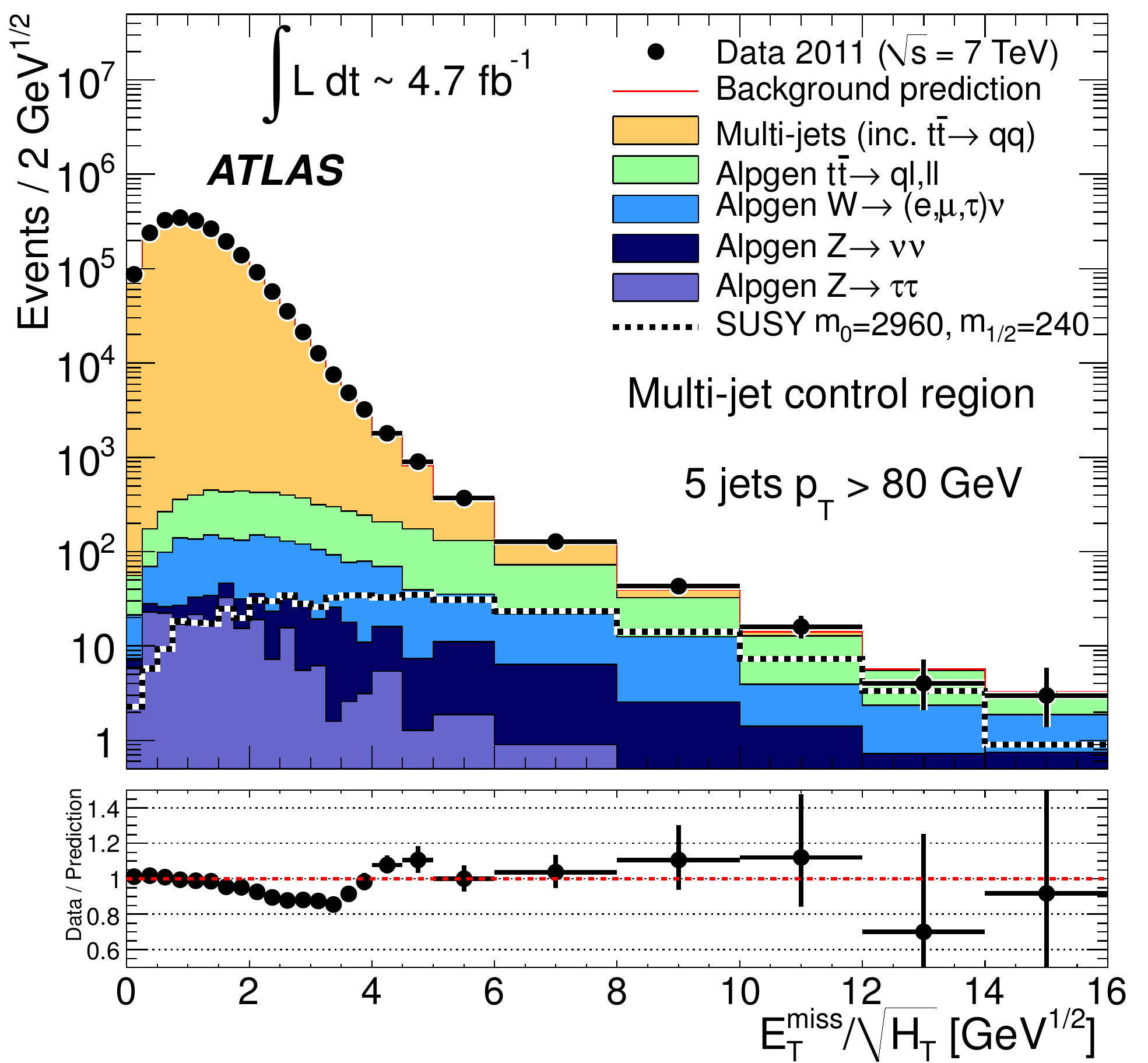} } \\
\caption{\label{fig:metsig_vr}
  \metsig{} distributions in example multi-jet control regions.
  \oursubref{fig:metsig_vr:55} For exactly six jets with $\pt{}>55\GeV$,
       compared to a prediction based on the \metsig{} distribution for exactly
       five jets with  $\pt{}>55\GeV$.
  \oursubref{fig:metsig_vr:80} For exactly five jets with $\pt{}>80\GeV$,
       compared to a prediction based on four jets with $\pt{}>80\GeV$.
   The multi-jet predictions have been normalized to the data
   in the region $\metsig<1.5\,\rootgev$ after subtraction of the predicted \leptonic{}
   backgrounds.
   The most important \leptonic{} backgrounds are also shown,
   based on MC simulations.
   Variable bin sizes are used with bin widths (in units of \rootgev{}) of
   0.25 (up to 4),
   0.5 (from 4 to 5),
   1 (from 5 to 6),
   and then 2 thereafter.
The error bars on the data points show the Poisson coverage interval corresponding to the number of data events observed in each bin.
}
\end{center}
\end{figure*}

The dominant background at intermediate values of 
\met{} is multi-jet production
including purely strong interaction processes and
fully hadronic decays of \ttbar.
These processes are not reliably predicted with existing Monte Carlo calculations, and so their contributions must be determined from collision data. 
Indeed, the selection cuts have been designed such that multi-jet processes can be determined reliably from supporting measurements.

The method for determining the multi-jet background from data is motivated by the following considerations.
In events dominated by jet activity, including hadronic decays of top quarks and gauge bosons,
the \met{} resolution is approximately proportional to $\sqrt{\HT}$, and is almost independent of the jet multiplicity. 
The distribution of the ratio $\metsig$ has a shape that is almost invariant under changes in the
jet multiplicity, as shown in \figref{fig:metsig_vr}.
The multi-jet backgrounds therefore can be determined using control regions with lower \metsig{} and/or lower jet multiplicity than the signal regions.\footnote{Residual variations in the shape of the \metsig{} are later used to quantify the systematic uncertainty associated with the method, as described in \oursecref{sec:backgrounds:qcd:systematic}.}
The control regions are assumed to be dominated by Standard Model processes, an assumption that is corroborated by the agreement of multi-jet cross section measurements with up to six jets~\cite{Aad:2011tqa} with Standard Model predictions.

As an example, the estimation of the background expected in the 8j55 signal region is obtained as follows.
A template describing the shape of the \metsig{} distribution is obtained from
those events that contain exactly six jets, using the same 55\,\GeV{} \pt{} threshold
as the target signal region.
That six-jet \metsig{} template is normalized to the number of eight-jet
events observed in the region $\metsig<1.5\rootgev$ after subtraction of the 
\leptonic{} background expectation.
The normalized template then provides a prediction for the multi-jet 
background for the 8j55 signal region for which $\metsig>4\rootgev{}$. 

A similar procedure is used for each of the signal regions, and can be summarized as follows. 
For each jet \pt{} threshold $\pthreshold \in\{55\GeV,\,80\GeV\}$,  
control regions are defined for different numbers $\mult$ of jets 
found above $\pthreshold$.
The number of events $N_{\pthreshold,\mult}(\smin{},\,\smax{})$ 
for which \metsig{} (in units of \rootgev{}) lies between $\smin{}$ and $\smax{}$ is determined,
and the predicted \leptonic{} contributions
$L_{\pthreshold,\mult}(\smin{},\,\smax{})$ subtracted 
\begin{equation*}
N_{\pthreshold,\mult}^{\slashed{L}}(\smin{},\,\smax{}) = N_{\pthreshold,\mult}(\smin{},\,\smax{})
                                                   - L_{\pthreshold,\mult}(\smin{},\,\smax{}) .
\end{equation*}
Transfer factors 
\begin{equation*}
T_{\pthreshold,\mult} = \frac{N_{\pthreshold,\mult}^{\slashed{L}}(4,\,\infty)} 
                      {N_{\pthreshold,\mult}^{\slashed{L}}(0,\,1.5)   }
\end{equation*}
connect regions with the same $\pthreshold$ and $\mult$ with different \metsig.
The multi-jet prediction for the signal region is found from the product of the
$T_{\pthreshold,\mult}$, with the same $\pthreshold$ as the signal region and $\mult=6$ when $\pthreshold=55\GeV$ ($\mult=5$ when $\pthreshold=80\GeV$) times the number of events (after subtracting the expected contribution from \leptonic{} background sources) satisfying signal region jet multiplicity requirements but with $\metsig<1.5\,\rootgev$.

\subsection{Systematic uncertainties on multi-jet backgrounds}\label{sec:backgrounds:qcd:systematic}

The method is validated by determining the accuracy of predictions 
for regions with jet multiplicities and/or \metsig{} smaller than those chosen for the signal regions.
\figref{fig:metsig_vr} shows that the shape of the \metsig{} distribution
for $\pthreshold=55\GeV$ and $\mult=6$ is predicted  
to an accuracy of better than 20\% from that measured using 
a template with the same value of $\pthreshold$ and $\mult=5$.
Similarly, the distribution for $\pthreshold=80\GeV$ and $\mult=5$ can be 
predicted for all \metsig{} using a template with $\mult=4$. 
The templates are normalized for $\metsig<1.5\,\rootgev$, 
and continue to provide a good prediction of the 
distribution out to values of $\metsig{}$ of $4\,\rootgev$ and beyond.
Additional validation regions are defined for each $\pthreshold$ 
and for jet multiplicity requirements equal to those
of the signal regions, but for the intermediate values of $(\smin{},\smax)$ of $(1.5,\,2)$, $(2,\,2.5)$ and $(2.5,\,3.5)$.
Residual inaccuracies in the predictions are used to quantify the systematic uncertainty from the closure of the method.
Those uncertainties are in the range 15\%--25\%, depending on \pthreshold{} and \metsig{}.

The mean number of
proton-proton interactions per bunch crossing $\langle\mu\rangle$
increased during the 2011 run, reaching $\langle\mu\rangle=16$
at the start of proton fills for runs late in the year. 
Sensitivity to those additional interactions is studied
by considering the jet multiplicity as a function of $\langle\mu\rangle$,
and of the number of reconstructed primary vertices.
The consistency of the high-\pt{} tracks within the selected jets
with a common primary vertex is also investigated.
The effect of additional jets from pile-up interactions is found to be
significant for low-\pt{} jets but 
small for jets with $\pt>45\,\GeV$, 
and negligible for the jet selection used for the SRs.

The presence of multiple in-time and out-of-time $pp$ interactions
also leads to a small but significant deterioration of the \met{} resolution.
The effectiveness of the \metsig{} template method described above
is tested separately for subsets of the data
with different values of the instantaneous luminosity, 
and hence of $\langle\mu\rangle$. 
Good agreement is found separately for each subset of the data.
Since the data set used to form the template has the same pile-up conditions
as that used to form the signal regions, 
the changing shape of the \met{} resolution is
included in the data-driven determination
and does not lead to any additional systematic uncertainty.

Due to the presence of neutrinos produced in the decay of hadrons containing
bottom or charm quarks, events with heavy-flavour jets exhibit a different 
\met{} distribution. To quantify the systematic uncertainty
associated with this difference, separate templates are defined
for events with at least one $b$-tagged jet and for those with none.
The sum of the predictions for events with and without $b$-tagged jets 
is compared to the flavour-blind approach, and the difference is used
to characterize the systematic uncertainty from heavy flavour (10\%--20\%). 
Other systematic uncertainties account for
imperfect knowledge of: 
the subtracted \leptonic{} contributions (10$\%$),
the potential trigger inefficiency (2\%),
and imperfect response of the calorimeter in problematic areas (1\%).

The backgrounds from multi-jet processes are cross checked using another
data-driven technique~\cite{Aad:2011ib} 
which smears the energies of individual jets from low-\met{} multi-jet `seed' events in data.
Separate smearing functions are defined for $b$-tagged and non-$b$-tagged
jets, with each modelling both the Gaussian core and the non-Gaussian tail of the jet response,
including the loss of energy from unobserved neutrinos.
The jet smearing functions are derived from {\tt GEANT4}~\cite{Agostinelli:2002hh} simulations~\cite{Aad:2010ah}. The Gaussian core of the function is tuned to di-jet data, and the non-Gaussian tails 
are verified with data in three-jet control regions in which the \ourvecptmiss{} can be associated
with the fluctuation of a particular jet. 
There is agreement within uncertainties between the background
predicted by this jet-smearing method and the primary method based on the
shape invariance of \metsig{}. 

\section{\Leptonic{} backgrounds}\label{sec:backgrounds:leptonic}

\begin{figure}
\begin{center}
\subfigure[\label{fig:sm_vr_top_45}]{\includegraphics[width=0.42\linewidth]{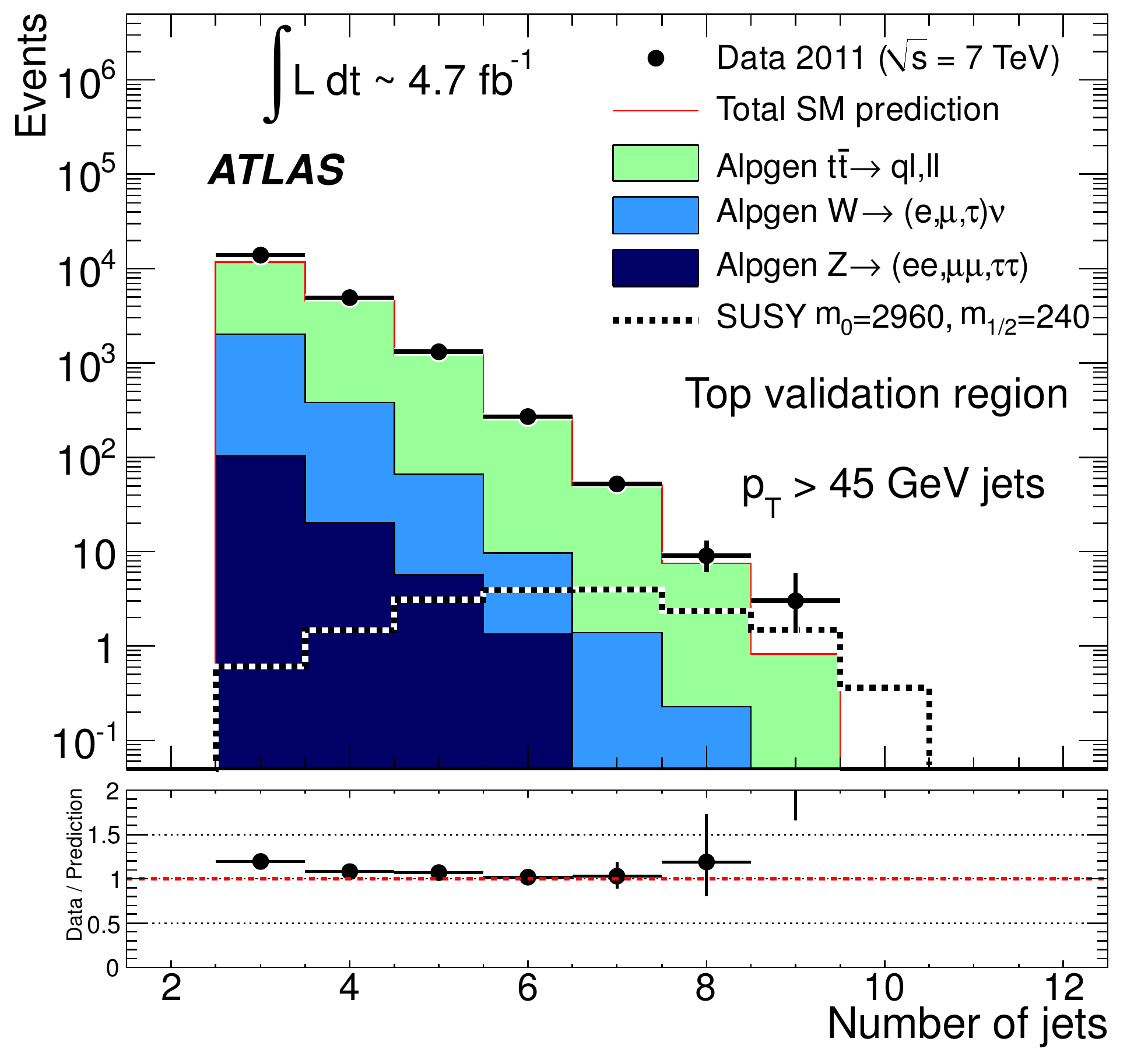}}
\subfigure[\label{fig:sm_cr_top_45}]{\includegraphics[width=0.42\linewidth]{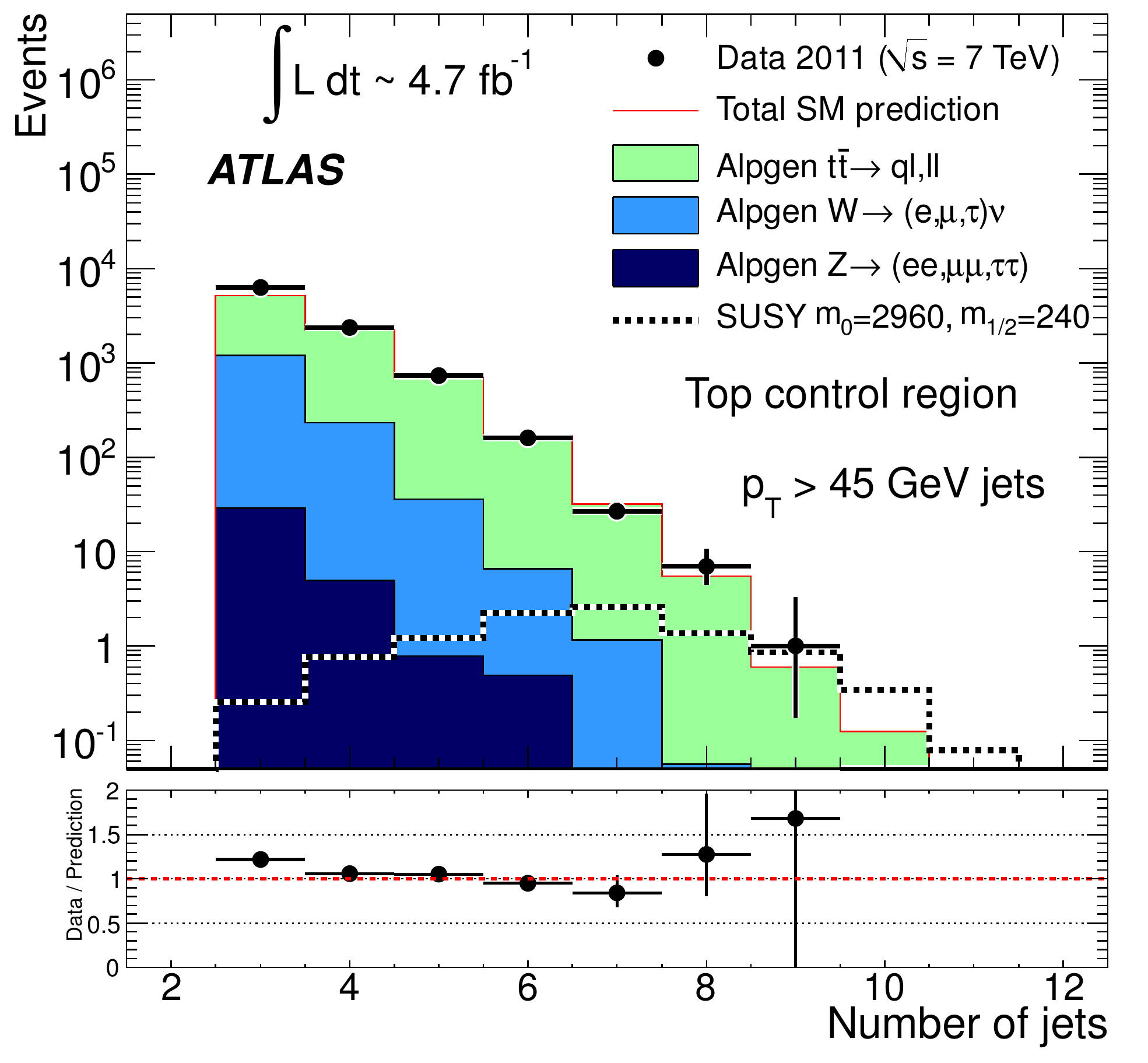}} \\

\subfigure[\label{fig:sm_vr_top_55}]{\includegraphics[width=0.42\linewidth]{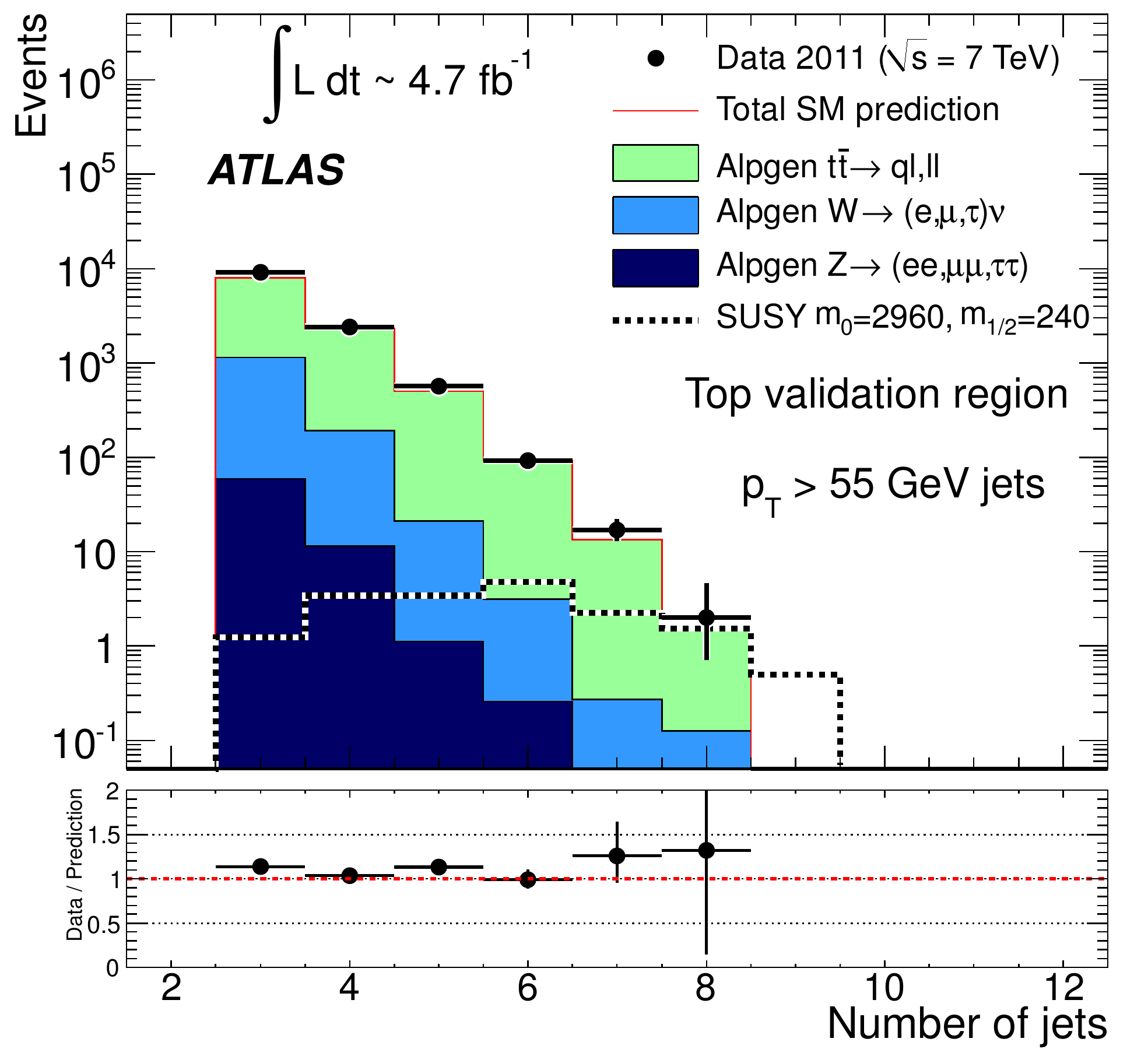}}
\subfigure[\label{fig:sm_cr_top_55}]{\includegraphics[width=0.42\linewidth]{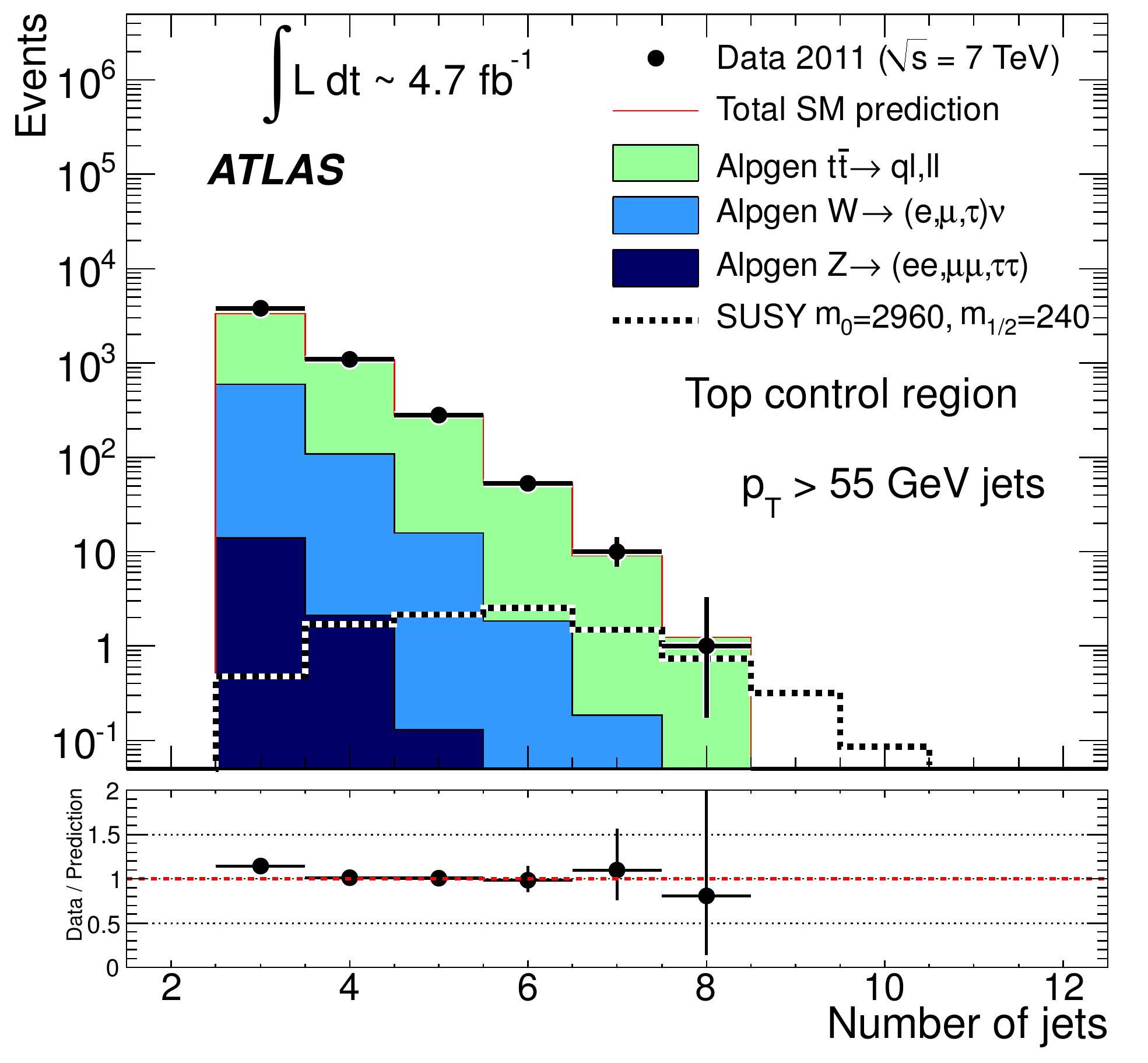}} \\
\subfigure[\label{fig:sm_vr_top_80}]{\includegraphics[width=0.42\linewidth]{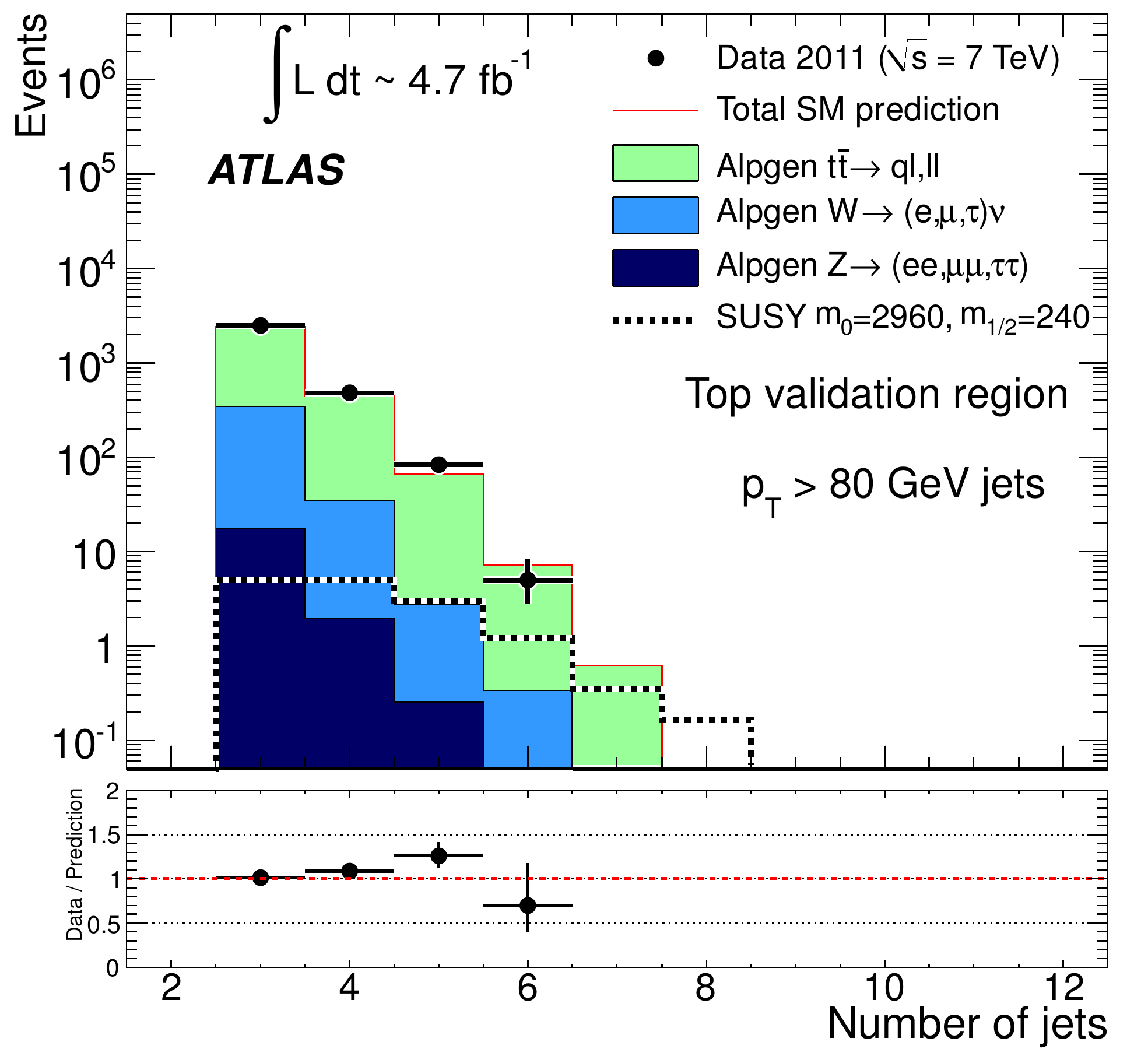}}
\subfigure[\label{fig:sm_cr_top_80}]{\includegraphics[width=0.42\linewidth]{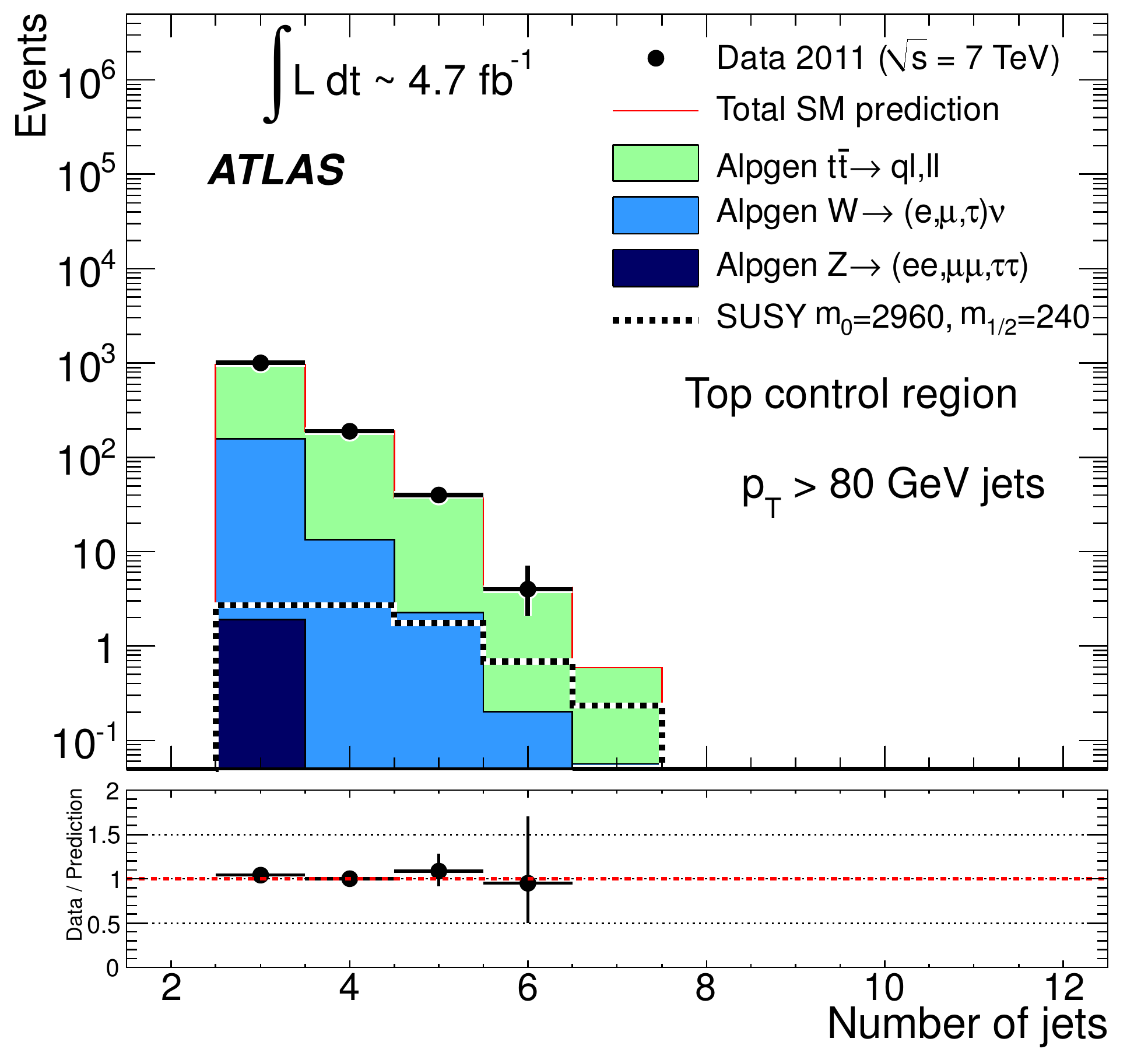}}
\caption{\label{fig:sm_vr_top}
  Jet multiplicity distributions for the $\ttbar$~+~jets 
  validation regions (left) 
  and control regions (right)
  before any jet multiplicity requirements, for a jet \pt{} threshold of $45\,\GeV{}$ (top), $55\,\GeV{}$ (middle) and $80\,\GeV{}$ (bottom).
}
\end{center}
\end{figure}

Non-fully-hadronic (i.e.~semi-leptonic or di-leptonic) $\ttbar$, and $W$ and $Z$ production are collectively referred to as \leptonic{} backgrounds.
The process $Z\rightarrow \nu\nu$~+~jets contributes to the signal regions since it produces jets in association with \met{}. 
Leptonic $\ttbar$ and $W$ decays contribute to the signal regions when hadronic $\tau$ decays allow them to evade the lepton veto, with smaller contributions from events in which electrons or muons are produced but are not reconstructed.

The \leptonic{} background predictions employ the Monte Carlo simulations described in \oursecref{sec:mc}.
To reduce uncertainties from Monte Carlo modelling and detector response, it is desirable to normalize the background predictions to data using control regions (CR) and cross-check them against data in other validation regions (VR). 
These control regions and validation regions are designed to be distinct from, but kinematically close to, the signal regions. Each is designed to provide enhanced sensitivity to a particular background process.

\begin{table}\centering
\renewcommand\arraystretch{1.2}
\begin{tabular}{| l || c | c | c |}
\hline
           & \hfill\ttbar{} + jets \hfill & \hfill $W$ + jets \hfill& $Z$ + jets \\ \hline\hline
Muon kinematics & \multicolumn{3}{| c |}{$\pT>20\GeV$, $|\eta|<2.4$} \\ \hline
Muon multiplicity &  \multicolumn{2}{| c |}{$=1$} & $=2$ \\ \hline
Electron multiplicity & \multicolumn{3}{| c |}{$=0$} \\ \hline
$b$-tagged jet multiplicity & $\geq 1$ & $=0$ & --- \\ \hline
\mT{} or $m_{\mu\mu}$ & \multicolumn{2}{| c |}{$50 \GeV < \mT < 100 \GeV$}
                                              & $80 \GeV < m_{\mu\mu} < 100 \GeV$ \\ \hline
\hline
${\rm VR}\rightarrow{\rm CR}$ transform & \multicolumn{2}{| c |}{ $\mu \rightarrow\textrm{jet}$} & $\mu \rightarrow \nu$ \\ \hline
\hline
Jet \pT{}, $|\eta|$, multiplicity (CR) &  \multicolumn{3}{| c |}{ \multirow{2}{*}{As in \tabref{tab:sr}.}}\\ 
\cline{1-1}
\metsig{} (CR) & \multicolumn{3}{| c |}{} \\ \hline
\end{tabular}
\caption{\label{tab:leptonic_cr}
Definitions of the validation regions and control regions 
for the \leptonic{} backgrounds: 
\ttbar{}~+~jets, $W$~+~jets and $Z$~+~jets.
The validation regions VR are defined by the first five selection requirements.
A long dash `---' indicates that no requirement is made.
The control regions CR differ from the VR 
in their treatment of the muons, and by having additional requirements on 
jets and \metsig{}, as shown in the final two rows.
}
\end{table}

The control and validation regions are defined as shown in~\tabref{tab:leptonic_cr}.
By using control regions that are kinematically similar to the signal regions,
theoretical uncertainties, including those arising from
the use of a leading-order (LO) generator, are reduced.
The $\ttbar$~+~jets and $W$~+~jets validation regions each require a single muon and no electrons. 
For the \ttbar{} process the single-muon selection is primarily sensitive to the semi-leptonic decay.\footnote{The procedure is also sensitive to those di-leptonic \ttbar{} decays in which one lepton was not observed in the VR. After the VR~$\rightarrow$~CR replacement $(\mu\rightarrow{\rm jet}$), the procedure captures the leading di-leptonic \ttbar{} contributions to the SR.}
The $\ttbar$~+~jets validation region is further enhanced by the requirement of at least one $b$-tagged jet, whereas for $W$~+~jets enhancement a $b$-tag veto is applied.
Since it is dominantly through hadronic $\tau$ decays that $W$ and $\ttbar$ contribute to the signal regions, the corresponding control regions are created by recasting the muon as a ($\tau$-)jet. 
For $Z\rightarrow \nu\nu$~+~jets the validation regions select events from the closely related process $Z\rightarrow \mu\mu$~+~jets.
The related control regions are formed from these validation regions by recasting the muons as neutrinos.

In detail, for those control regions where the Monte Carlo simulations predict at least one event for \ourintlumi{}, 
the leptonic background prediction $s_i$ for each signal region from each background is calculated by 
multiplying the number of data events $c_i^{\rm data}$  found in the corresponding control region 
by a Monte Carlo-based factor $t_{i}^{\rm MC}$ 
\[
s_i = c_i^{\rm data} \times t_{i}^{\rm MC}.
\]
This transfer factor is defined to be the ratio of the 
number of MC events found in the signal region to the number of MC events found in the control region 
\[
t_i^{\rm MC} = \frac{s_i^{\rm MC}}{c_i^{\rm MC}}.
\]
In each case, the event counts are corrected for the expected contamination by the other background processes.
Whenever less than one event is predicted in the control region, 
the Monte Carlo prediction for the corresponding signal region is used directly, without 
invoking a transfer factor.

For the $\ttbar$~+~jets background, 
the validation region requires exactly one isolated muon, 
at least one $b$-tagged jet,
and no selected electrons.
The transverse mass for the muon 
transverse momentum $\vec{p}_{\rm T}^{\,\mu}$ and the missing transverse
momentum two-vector $\ourvecptmiss$ is calculated using 
massless two-vectors
\begin{equation*}
\mT^2 = 2 |\vec{p}_{\rm T}^{\,\mu}||\ourvecptmiss| - 2 \vec{p}_{\rm T}^{\,\mu} \cdot \ourvecptmiss ~,
\end{equation*}
and must satisfy $50\,\GeV < \mT < 100\,\GeV$.
\figref{fig:sm_vr_top} shows the jet multiplicity in the \ttbar{}
validation regions, and it is demonstrated that the Monte Carlo provides a good description of the data. 

The $\ttbar$ control regions used to calculate the background expectation
differ from the validation regions as follows.
Since the dominant source of background is from hadronic $\tau$ decays
in the control regions, the muon is used to mimic a jet, as follows.
If the muon has sufficient \pT{} to pass the jet selection threshold $\pthreshold$, the jet multiplicity
is incremented by one. If the muon \pT{} is larger than 40\,\GeV{} it is added to \HT{}.
The selection variable \metsig{} is then recalculated, and 
required to be larger than the threshold value of $4\,\rootgev$. 
Distributions of the jet multiplicity in the \ttbar{}
control regions may also be found in \figref{fig:sm_vr_top}.

\begin{figure*}
  \centering
  \subfigure[\label{fig:sm_vr_w:55}]{\includegraphics[width=0.45\linewidth]{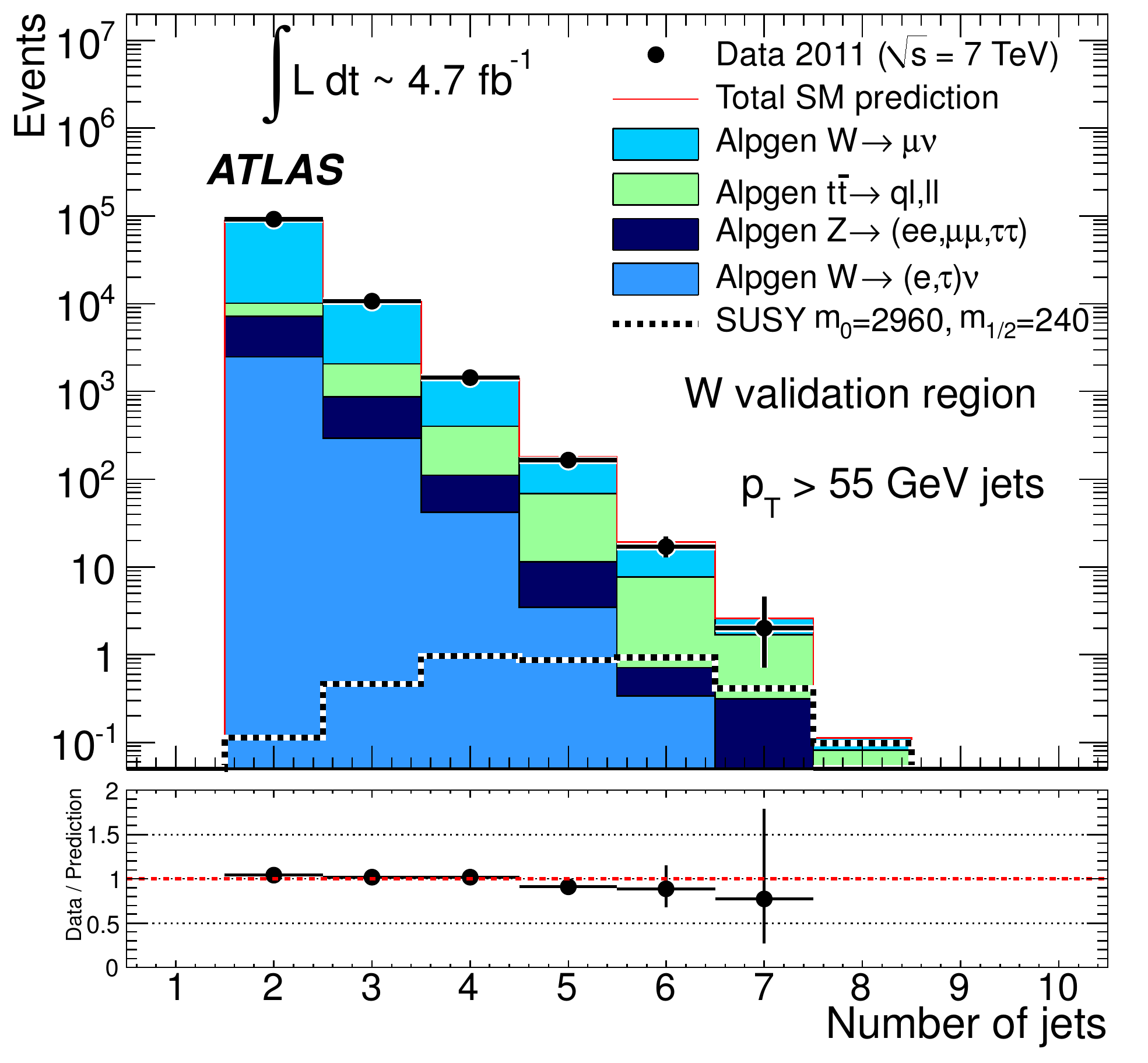} }
  \subfigure[\label{fig:sm_cr_w:55}]{\includegraphics[width=0.45\linewidth]{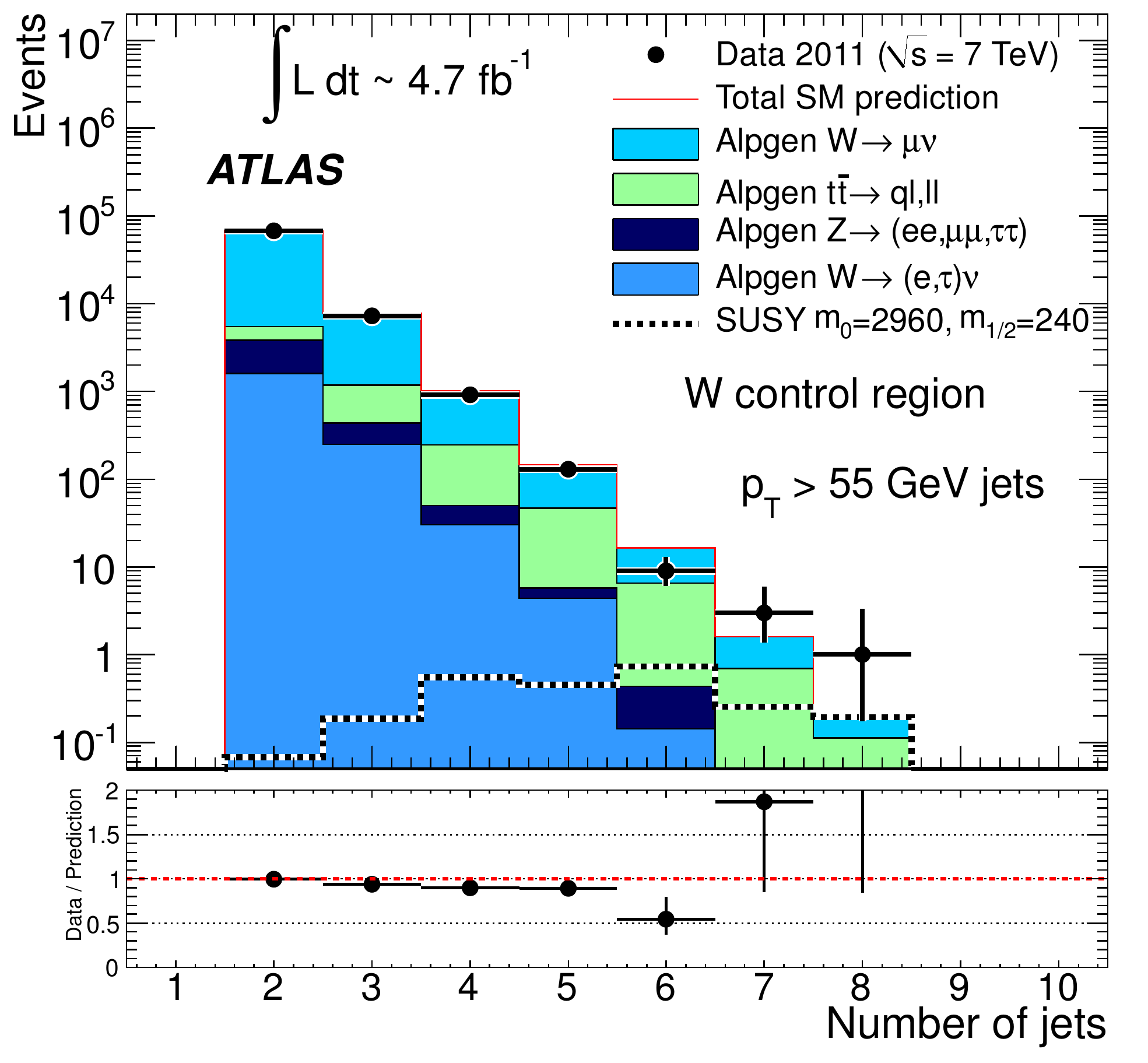} }\\
  \subfigure[\label{fig:sm_vr_w:80}]{\includegraphics[width=0.45\linewidth]{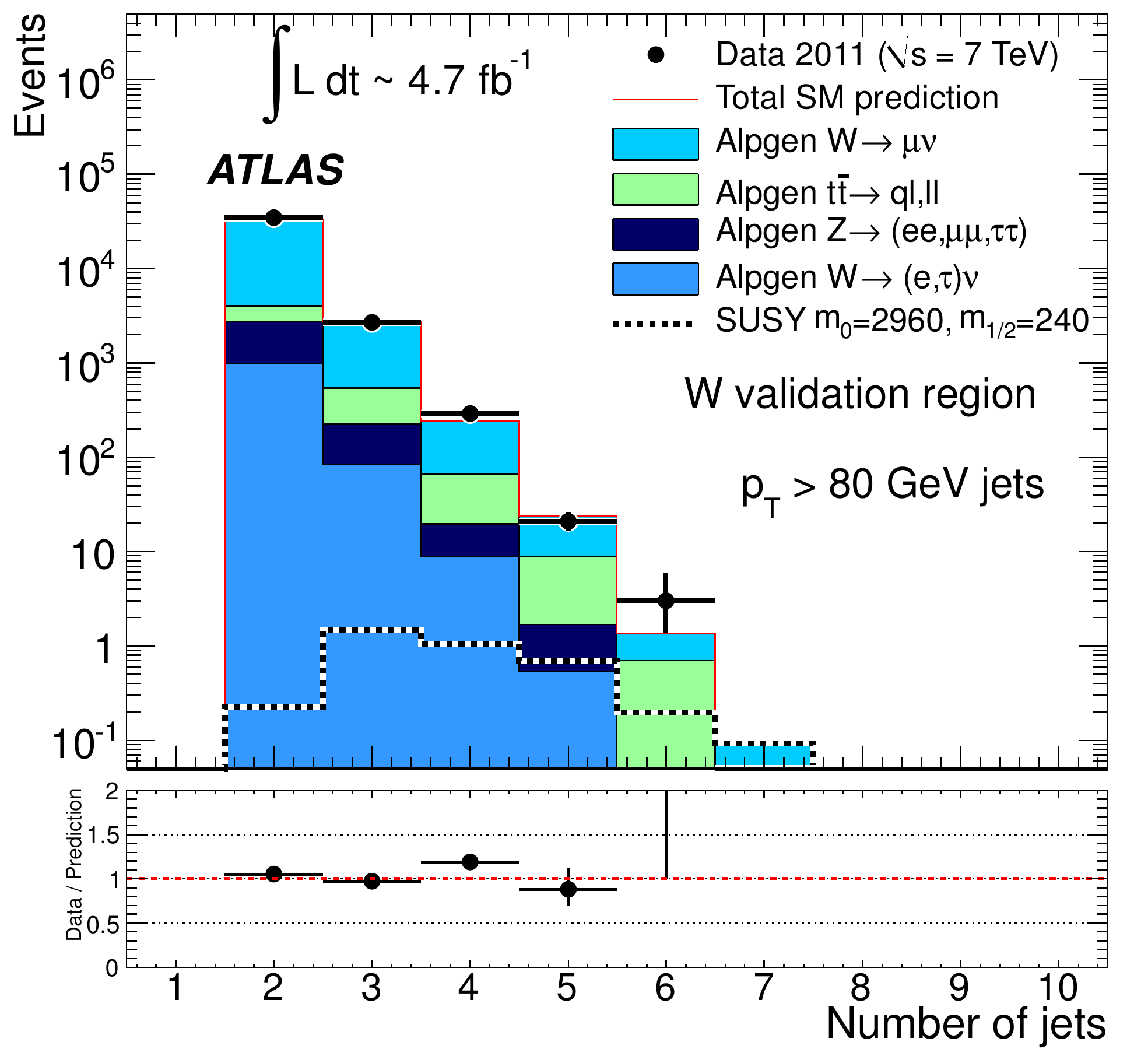} } 
  \subfigure[\label{fig:sm_cr_w:80}]{\includegraphics[width=0.45\linewidth]{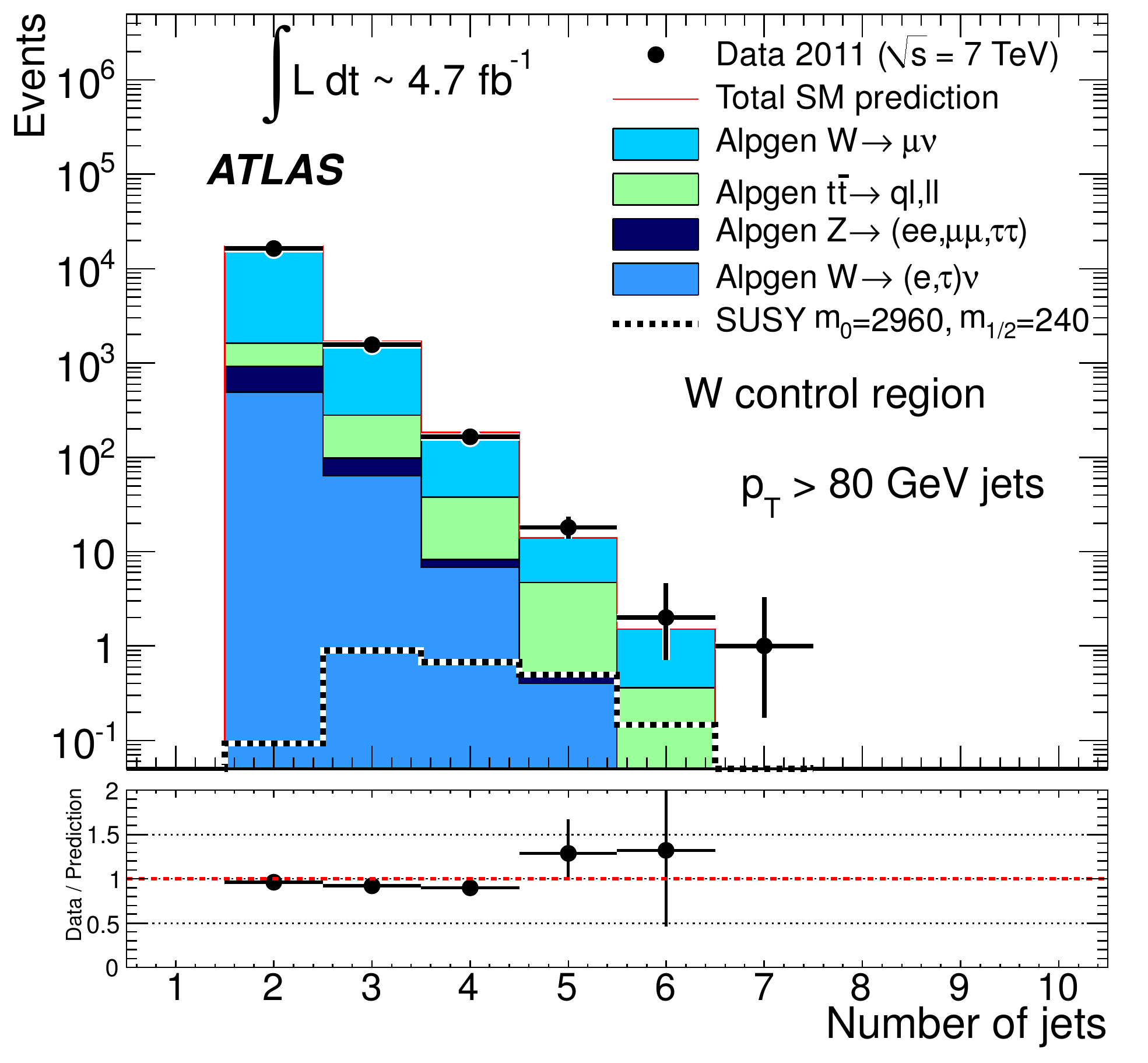} } 
  \caption{\label{fig:sm_cr_w}
    Jet multiplicity distributions for the $W^\pm$~+~jets validation regions
    (left) 
    and control regions (right)
    before any jet multiplicity requirements, and for a jet \pt{} threshold of $55\,\GeV{}$ (top) and $80\,\GeV{}$ (bottom).
  }
\end{figure*}

\begin{figure*}
  \centering
  \subfigure[\label{fig:sm_vr_z:55}]{\includegraphics[width=0.45\linewidth]{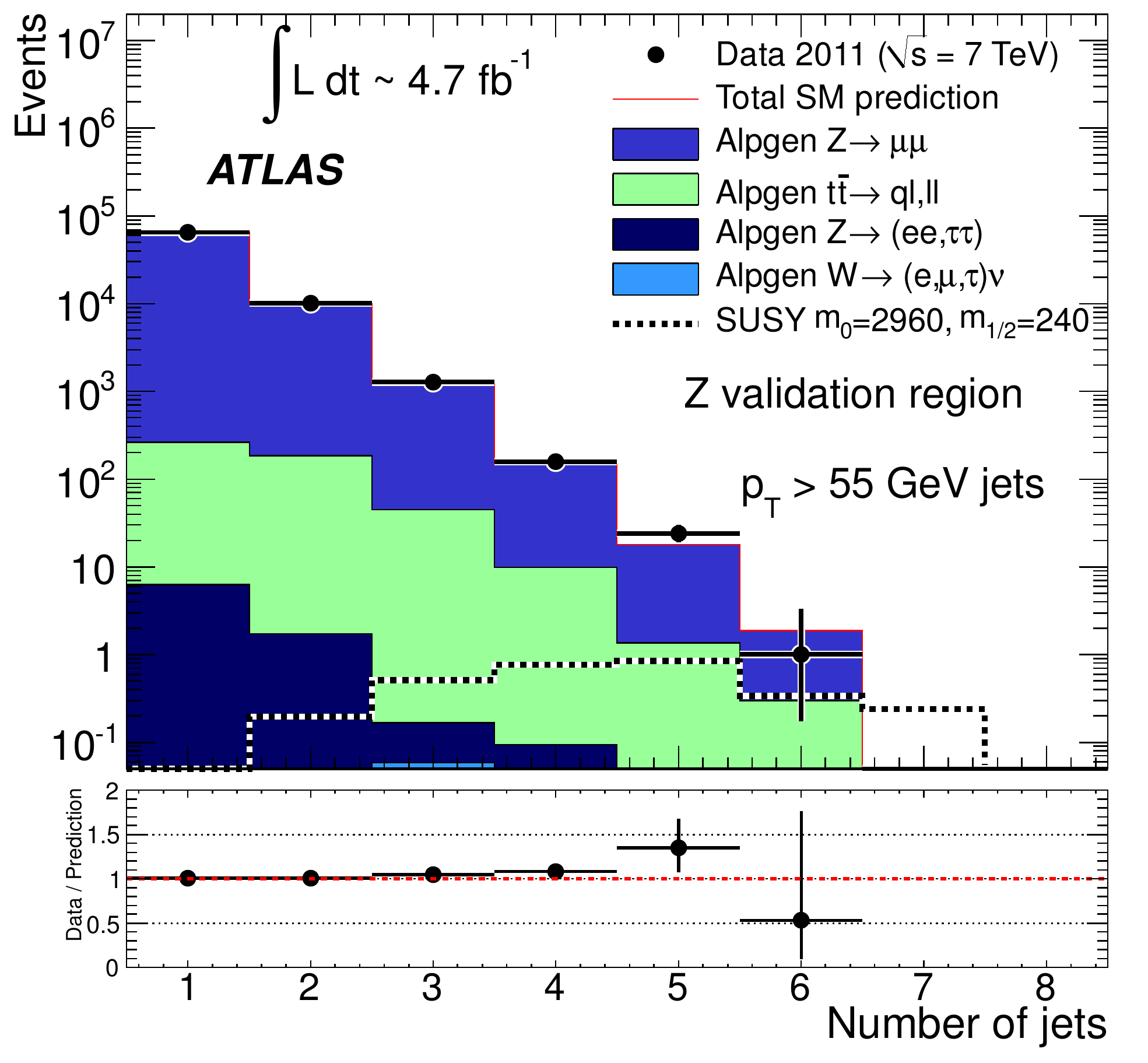} }
  \subfigure[\label{fig:sm_cr_z:55}]{\includegraphics[width=0.45\linewidth]{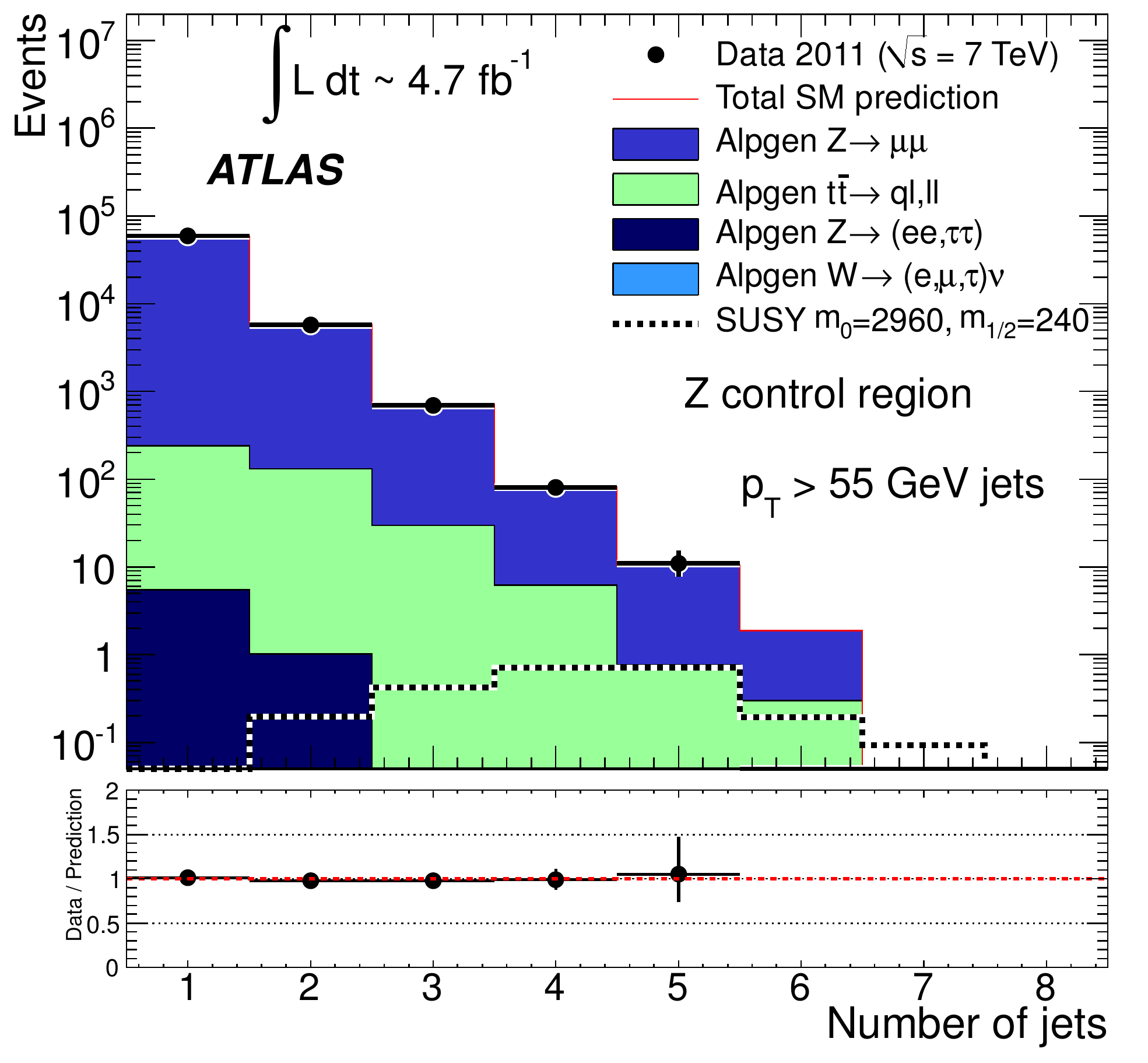} } \\
  \subfigure[\label{fig:sm_vr_z:80}]{\includegraphics[width=0.45\linewidth]{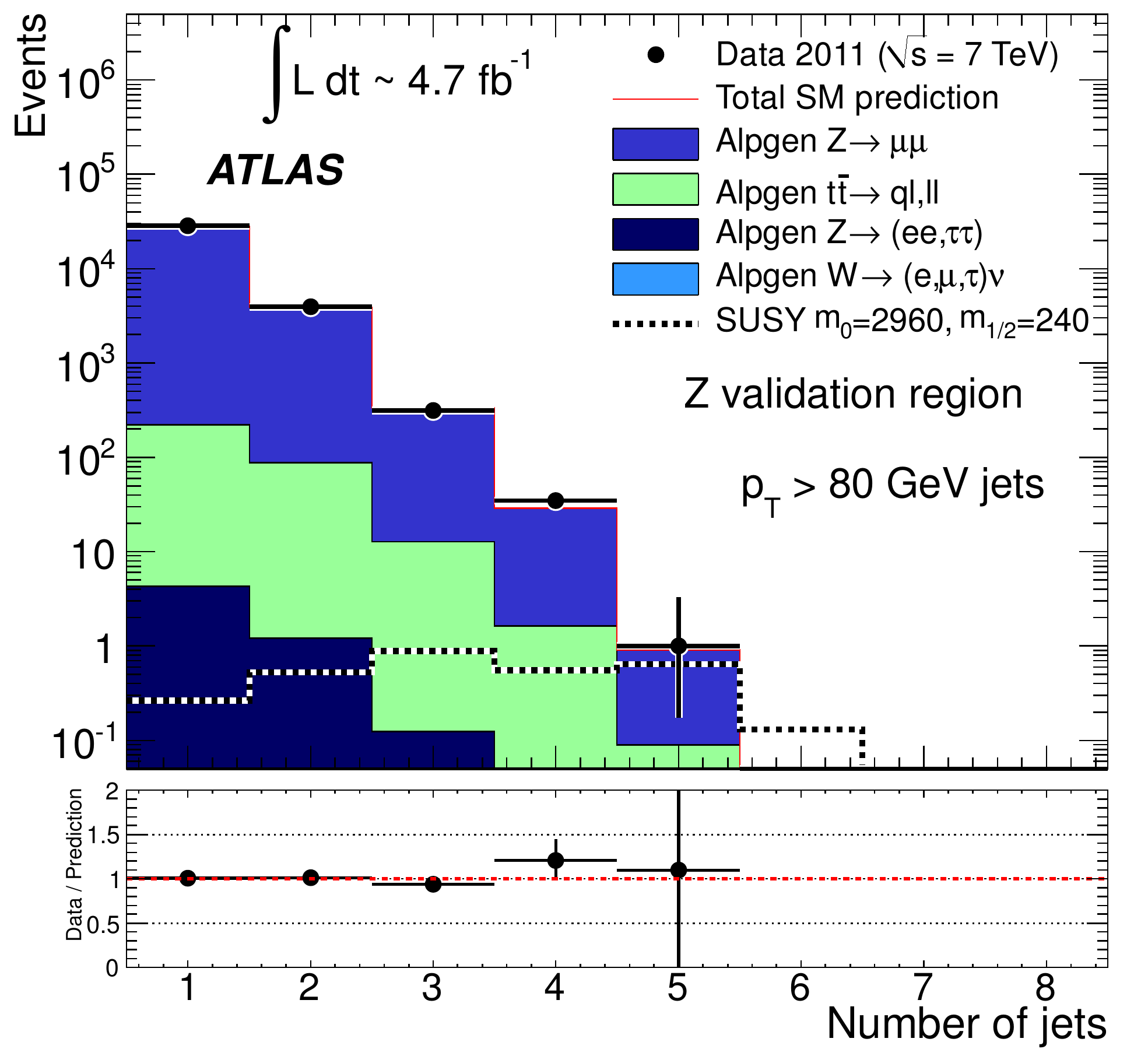} } 
  \subfigure[\label{fig:sm_cr_z:80}]{\includegraphics[width=0.45\linewidth]{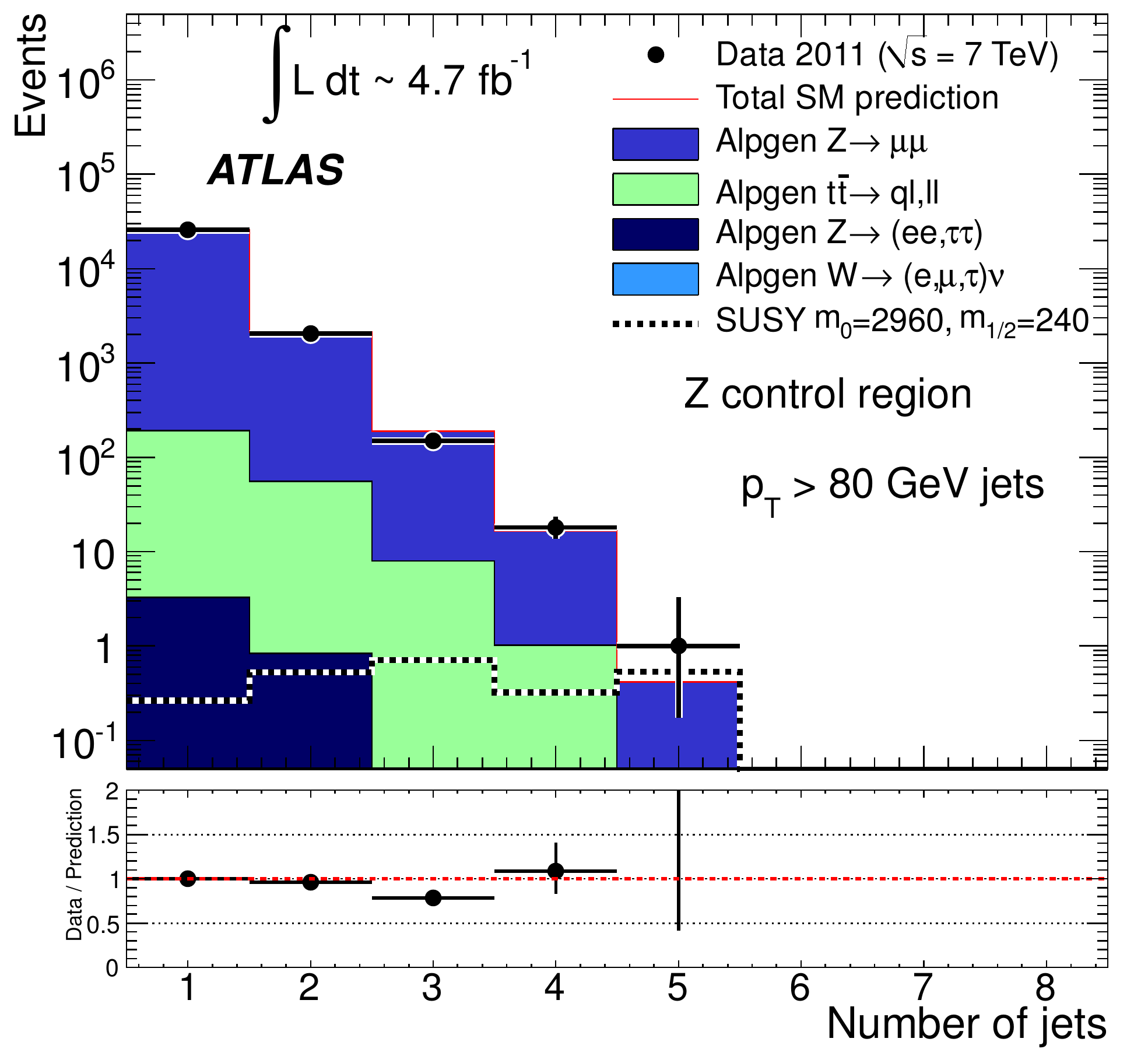} }
  \caption{\label{fig:sm_cr_z}
    As for \figref{fig:sm_cr_w} but for the $Z$~+~jets validation regions and control regions.
  }
\end{figure*}

The $W$~+~jets validation regions and control regions are 
defined in a similar manner to those for $\ttbar$~+~jets, 
except that a $b$-jet veto is used rather than a $b$-jet requirement
(see \tabref{tab:leptonic_cr}).
\figref{fig:sm_cr_w} shows that the resulting jet multiplicity
distributions are well described by the Monte Carlo simulations.
\par
The $Z$~+~jets validation regions are defined
(as shown in \tabref{tab:leptonic_cr})
requiring precisely two muons with invariant mass 
$m_{\mu\mu}$ consistent with $m_Z$.
The dominant backgrounds from $Z$~+~jets arise from
decays to neutrinos, so in forming the $Z$~+~jets control regions
from the validation regions,
the vector sum of the $\vec{p}_{\rm T}$ of the muons 
is added to the measured \ourvecptmiss, to model the \met{} expected
from $Z\rightarrow \nu\nu$ events. 
The selection variable \metsig{} is then recalculated
and required to be greater than $4 \rootgev$ for events in the control region.
\figref{fig:sm_cr_z} shows that the resulting jet multiplicity
distributions in both validation and control regions 
are well described by the Monte Carlo simulations.

For each of the \leptonic{} backgrounds further comparisons 
are made between Monte Carlo and data using the lower jet \pT{} threshold of $45\,\GeV$, 
showing agreement within uncertainties for all multiplicities 
(up to nine jets for $\ttbar$, see Figures~\ref{fig:sm_vr_top}(a) and \ref{fig:sm_vr_top}(b). 
The \alpgen{} Monte Carlo predictions for $Z$~+~jets and $W$~+~jets were 
determined with six additional partons in the matrix element calculation,
and cross checked with a calculation in which only five additional partons
were produced in the matrix element -- in each case with 
additional jets being produced in the parton shower. 
The two predictions are consistent with each other and with the data, 
providing further supporting evidence that the parton shower offers a
sufficiently accurate description of the additional jets.

\subsection{Systematic uncertainties on \leptonic{} backgrounds}

The use of control regions is effective in reducing uncertainties from Monte Carlo modelling and detector response.
When predictions are taken directly from the Monte Carlo, 
the \leptonic{} background determinations are subject to 
systematic uncertainties
from Monte Carlo modelling of: the jet energy scale (JES, $40\%$), 
the jet energy resolution (JER, $4\%$),
the number of multiple proton-proton interactions ($3\%$),
the $b$-tagging efficiency ($5\%$ for $\ttbar$), 
the muon trigger and reconstruction efficiency 
and the muon momentum scale.
The numbers in parentheses indicate the typical 
values of the SR event yield uncertainties prior to the partial cancellations 
that result from the use of control regions.

The JES and JER uncertainties are calculated using a combination of data-driven and
Monte Carlo techniques~\cite{Aad:2011he}, using the complete 2011 ATLAS data set.
The calculation accounts for the variation in the uncertainty
with jet \pt{} and $\eta$, and that due to nearby jets.
The Monte Carlo simulations model the multiple proton-proton interactions
with a varying value of $\langle\mu\rangle$ which is well matched
to that in the data.
The residual uncertainty from pile-up interactions is determined
by reweighting the Monte Carlo samples so that $\langle\mu\rangle$ is
increased or decreased by 10\%.
The uncertainty in the integrated luminosity is $3.9\%$~\cite{lumi_summer_11}.

When transfer factors are used to connect control regions to signal regions,
the effects of these uncertainties largely cancel in the ratio.
For example, the impact of the jet energy scale uncertainty is reduced to $\approx6\%$.

\section{Results, interpretation and limits}\label{sec:results}

\begin{figure*}
\centering
\subfigure[\label{fig:metsig:7j55} 7j55]{\includegraphics[width=0.42\linewidth]{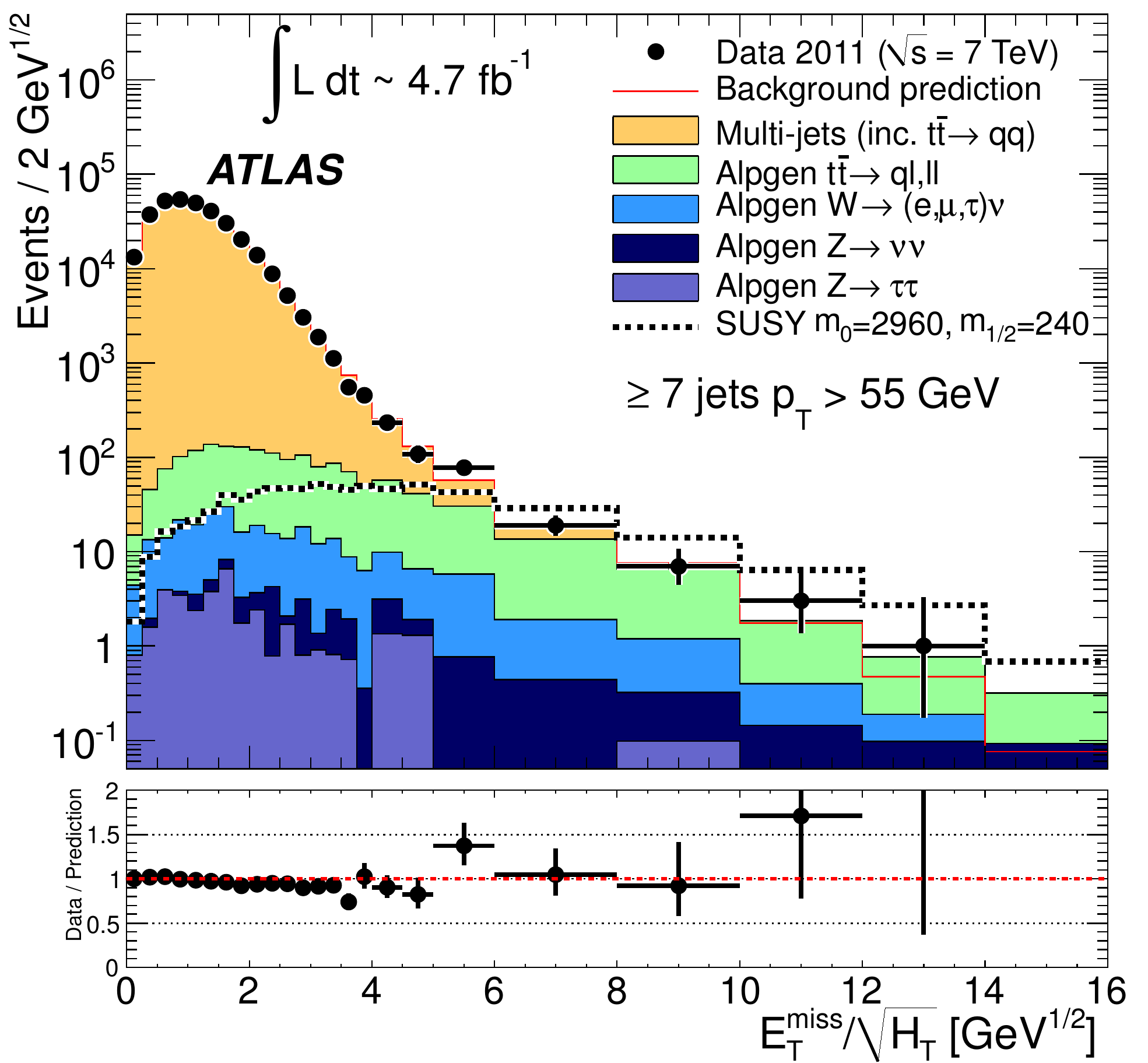} }
\subfigure[\label{fig:metsig:6j80} 6j80]{\includegraphics[width=0.42\linewidth]{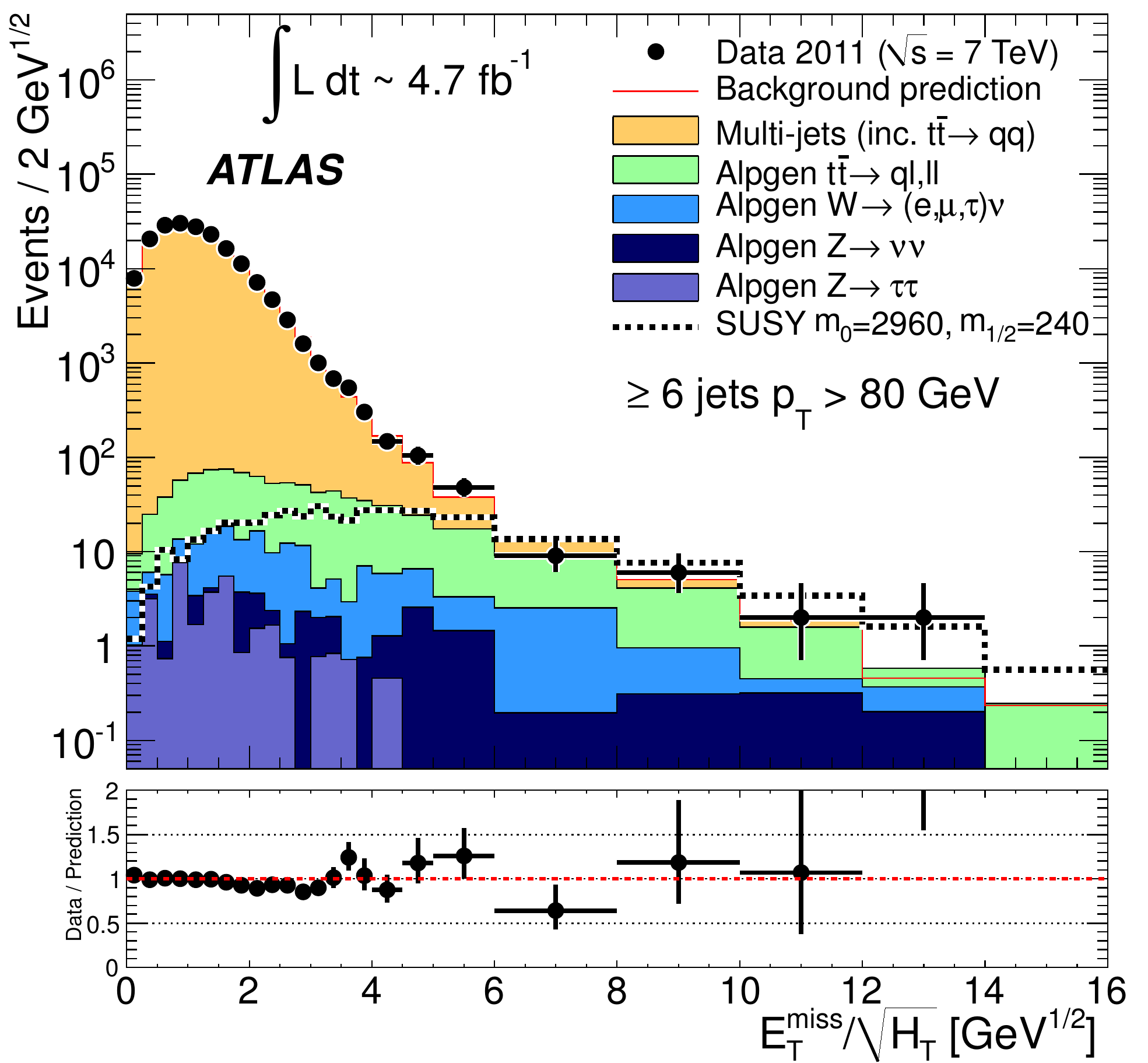} } \\
\subfigure[\label{fig:metsig:8j55} 8j55]{\includegraphics[width=0.42\linewidth]{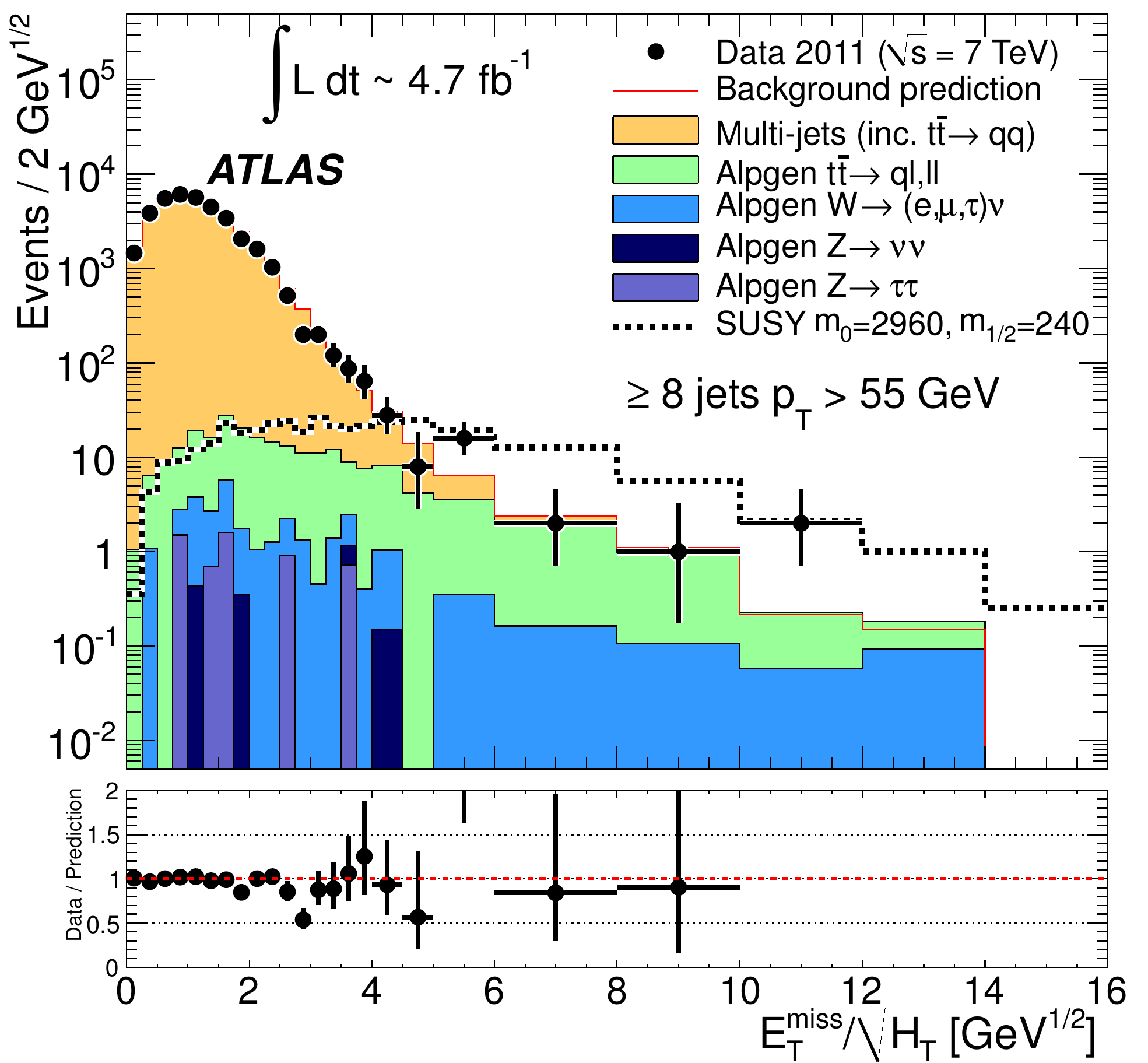} }
\subfigure[\label{fig:metsig:7j80} 7j80]{\includegraphics[width=0.42\linewidth]{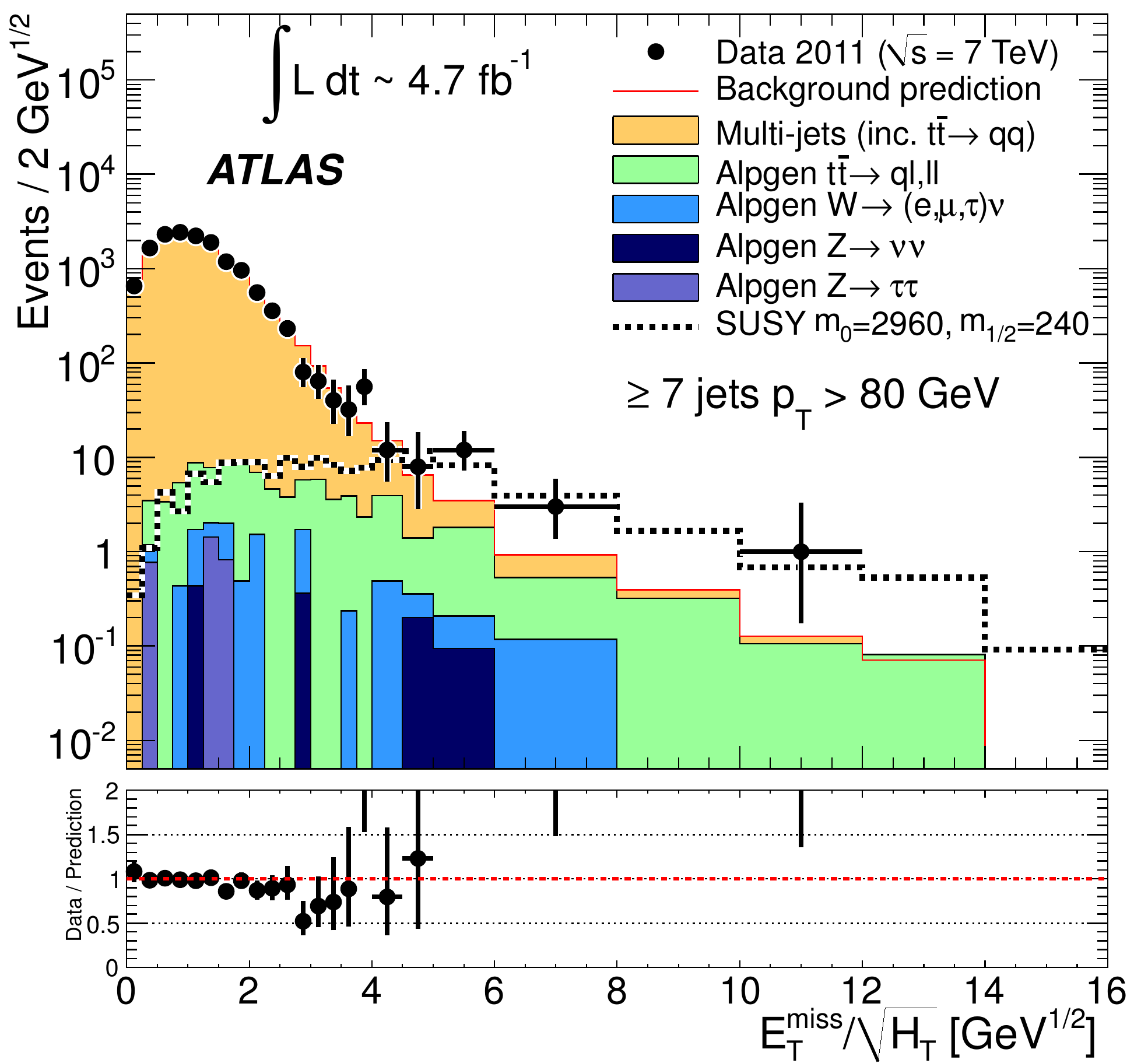} } \\
\subfigure[\label{fig:metsig:9j55} 9j55]{\includegraphics[width=0.42\linewidth]{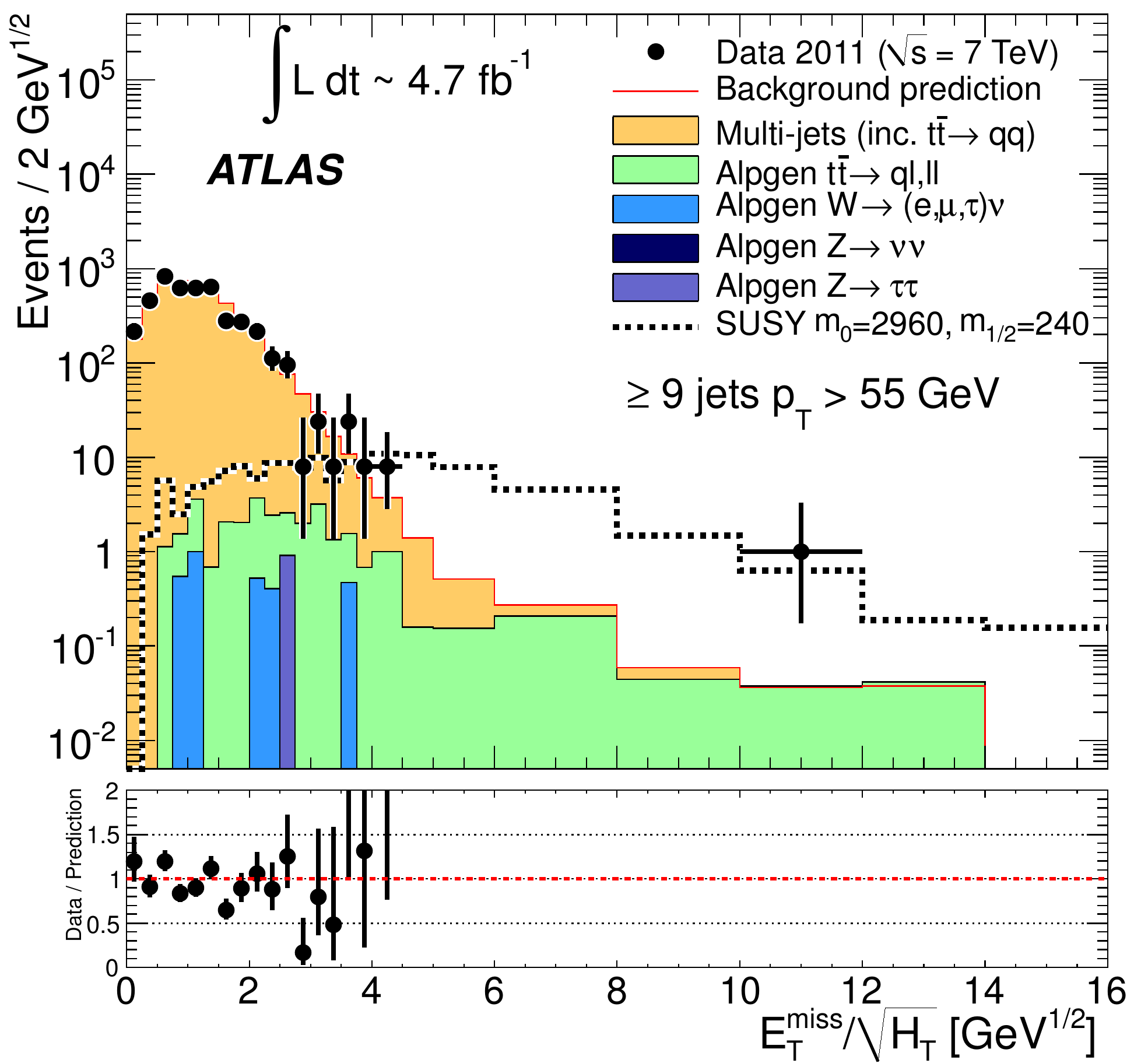} }
\subfigure[\label{fig:metsig:8j80} 8j80]{\includegraphics[width=0.42\linewidth]{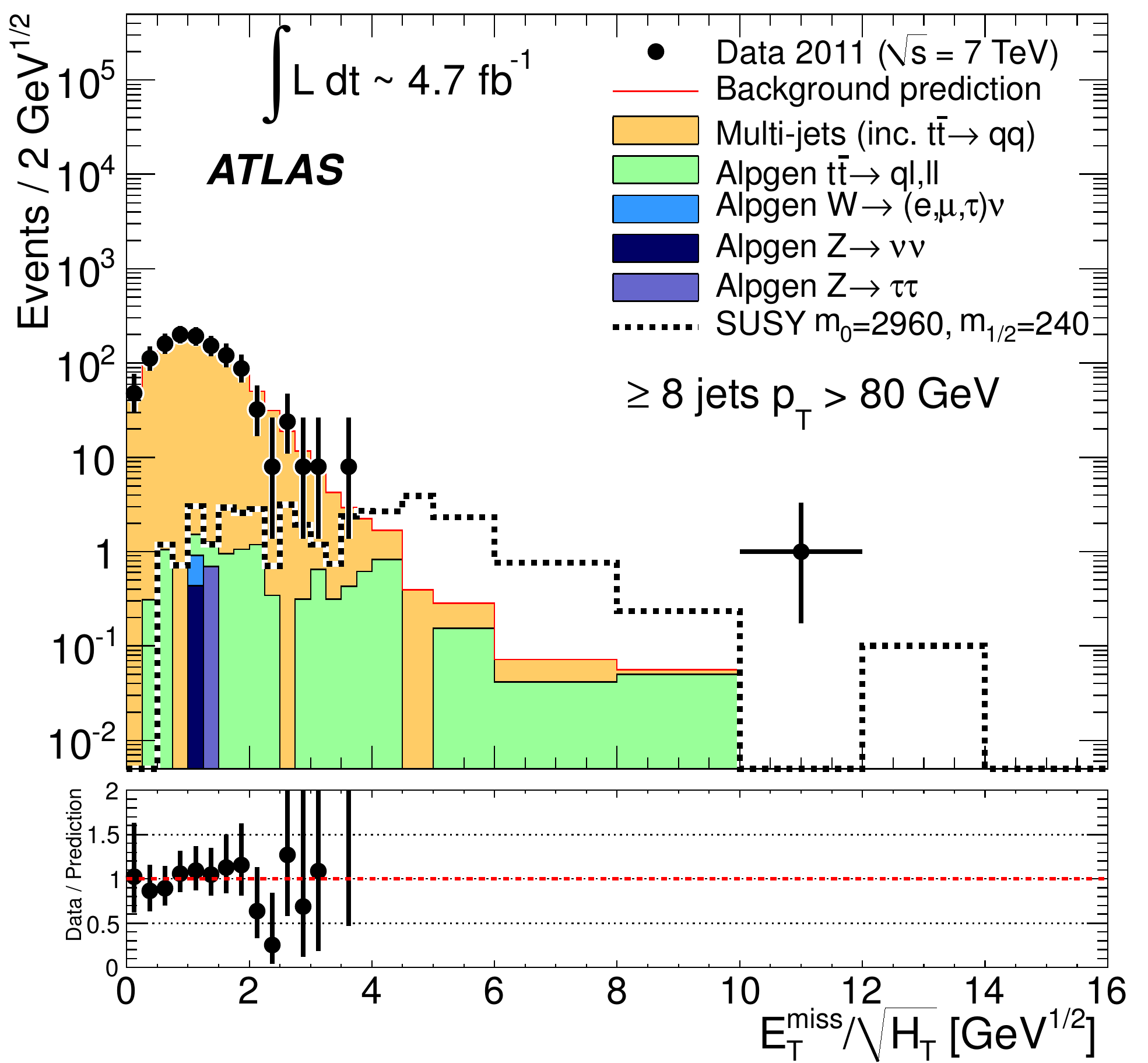} } \\
\caption{\label{fig:metsig}
  The distribution of the variable $\met{}/\sqrt{H_T}\,$ 
  for each of the six different signal regions defined in \tabref{tab:sr},
  prior to the final $\metsig{}>4\rootgev$ requirement. 
}
\end{figure*}

\begin{figure*}
\centering
\subfigure[\label{fig:mult:55}]{\includegraphics[width=0.65\linewidth]{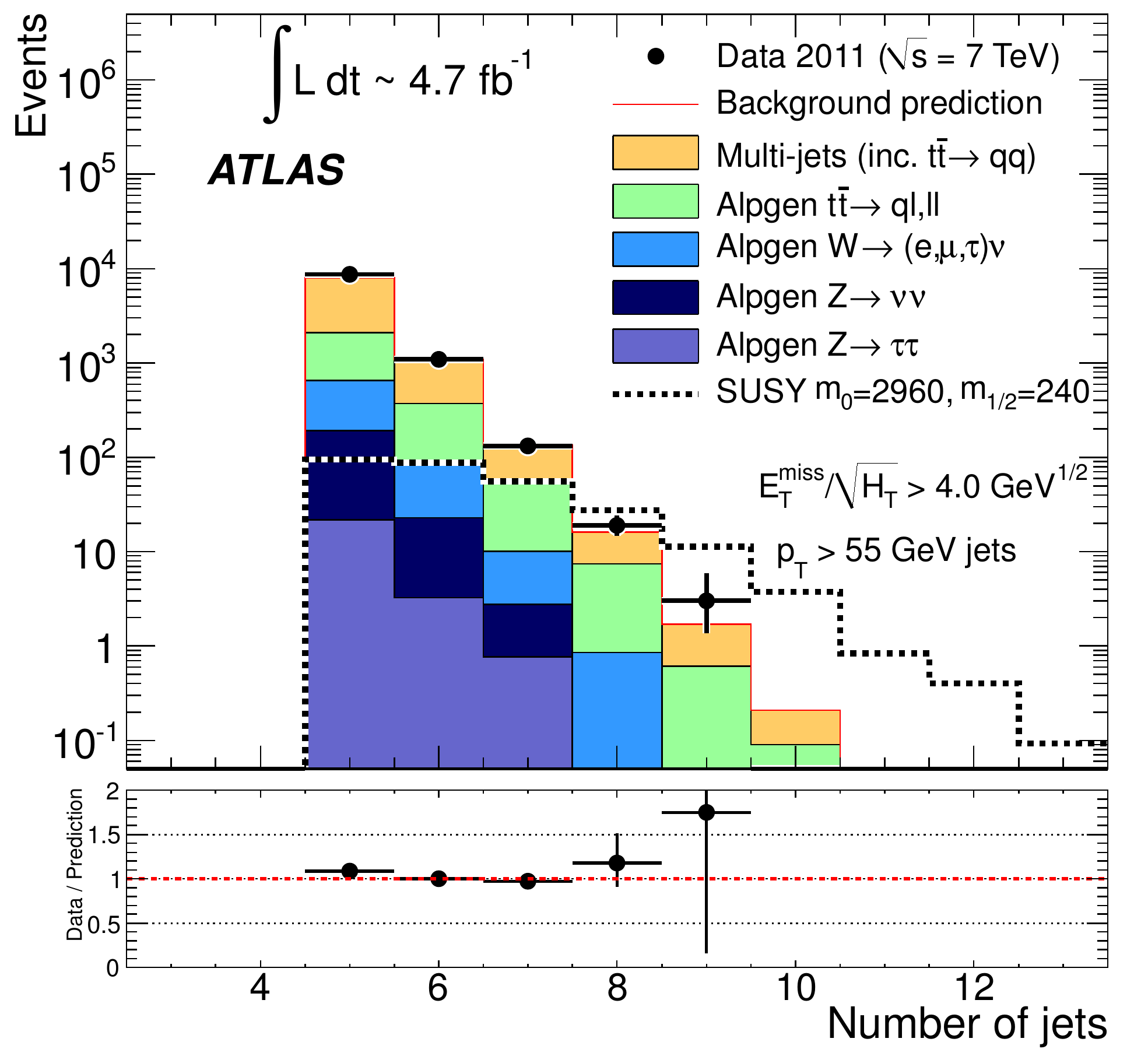} }
\subfigure[\label{fig:mult:80}]{\includegraphics[width=0.65\linewidth]{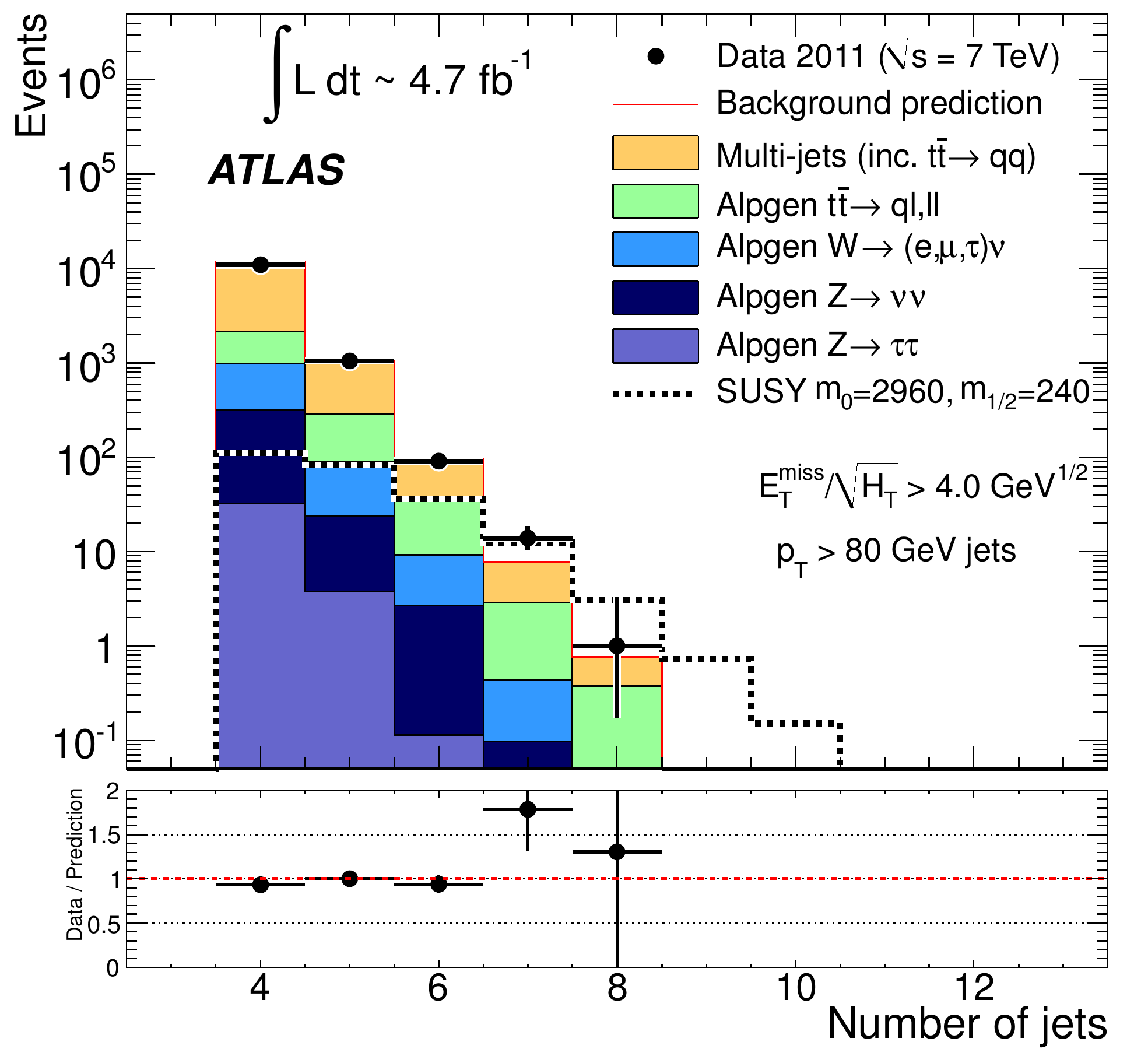} } \\
\caption{\label{fig:jetmult_signal}
  The distribution of jet multiplicity for jets with $\pT{}>55\,\GeV$ \oursubref{fig:mult:55} and those with $\pt{}>80\,\GeV$ \oursubref{fig:mult:80}.
  Only events with $\metsig > 4 \rootgev$ are shown.
}
\end{figure*}

\begin{table*}
\renewcommand\arraystretch{1.25}
\footnotesize
\begin{center}
\begin{tabular}{ | l || c | c | c | c | c | c |  } 
\hline
{\bf Signal region} & {\bf ~7j55~} & {\bf ~8j55~} & {\bf ~9j55~} & {\bf ~6j80~} & {\bf ~7j80~} & {\bf ~8j80~} \\ \hline\hline
Multi-jets & 91$\pm$20 & 10$\pm$3 & 1.2$\pm$0.4 &
67$\pm$12 & 5.4$\pm$1.7 & 0.42$\pm$0.16  \\ \hline
$\ttbar{} \rightarrow q\ell,\ell\ell$ & 55$\pm$18 & 5.7$\pm$6.0 & 0.70$\pm$0.72
& 24$\pm$13 & 2.8$\pm$1.8 & 0.38$\pm$0.40   \\ \hline
$W$ + jets  & 18$\pm$11 & 0.81$\pm$0.72 & 0+0.13 & 13$\pm$10 & 0.34$\pm$0.21 &
0+0.06 \\ \hline
$Z$ + jets & 2.7$\pm$1.6 & 0.05$\pm$0.19 & 0+0.12 & 2.7$\pm$2.9 & 0.10$\pm$0.17 & 0+0.13   \\ \hline\hline
{\bf Total Standard Model } &
\bf 167$\pm$34 & 
\bf 17$\pm$7 & 
\bf 1.9$\pm$0.8 & 
\bf 107$\pm$21 & 
\bf 8.6$\pm$2.5 & 
\bf 0.80$\pm$0.45  \\ \hline\hline
{\bf Data  }  &
\bf 154 &
\bf 22 &
\bf 3 &
\bf 106 &
\bf 15 &
\bf 1 \\ \hline\hline
\nmaxbsm{} (exp) &
72 &
16 &
4.5 &
46 &
8.4 &
3.5 \\
\nmaxbsm{} (obs) &
64 &
20 &
5.7 &
46 &
15 &
3.8            \\ \hline
$\sigmamaxbsm \cdot A \cdot \epsilon$ (exp) [fb] &
15 &
3.4 &
0.96 &
9.8 &
1.8 &
0.74 \\
$\sigmamaxbsm \cdot A \cdot \epsilon$ (obs) [fb]&
14 &
4.2 &
1.2 &
9.8 &
3.2 &
0.81              \\ \hline
$p_{\rm SM}$ &
0.64 &
0.27 &
0.28 &
0.52 &
0.07 &
0.43       \\ \hline
\end{tabular}
\caption{\label{tab:results}
Results for each of the six signal regions for an integrated luminosity of \ourintlumi{}.
The expected numbers of Standard Model events are given for
each of the following sources:
multi-jet (including fully hadronic \ttbar{}),
semi- and fully-leptonic $\ttbar$ decays combined,
and $W$ and $Z$ bosons (separately) in association with jets,
as well as the total Standard Model expectation.
The uncertainties on the predictions show the combination of the statistical and systematic components.
Where small event counts in control regions have not made it possible to determine
a central value for the expectation, an asymmetric bound is given instead.
The numbers of observed events are also shown.
The final five rows show the statistical quantities described in the text.
Both the expected (exp) and the observed (obs) values are shown for
$\nmaxbsm$ and $\sigmamaxbsm \times A \times \epsilon$. 
}
\end{center}
\end{table*}

\begin{figure*}
\vspace{-15mm}
\centering
\subfigure[\label{fig:exclusion_combined:msugra} \msugra ]{\includegraphics[width=0.62\linewidth]{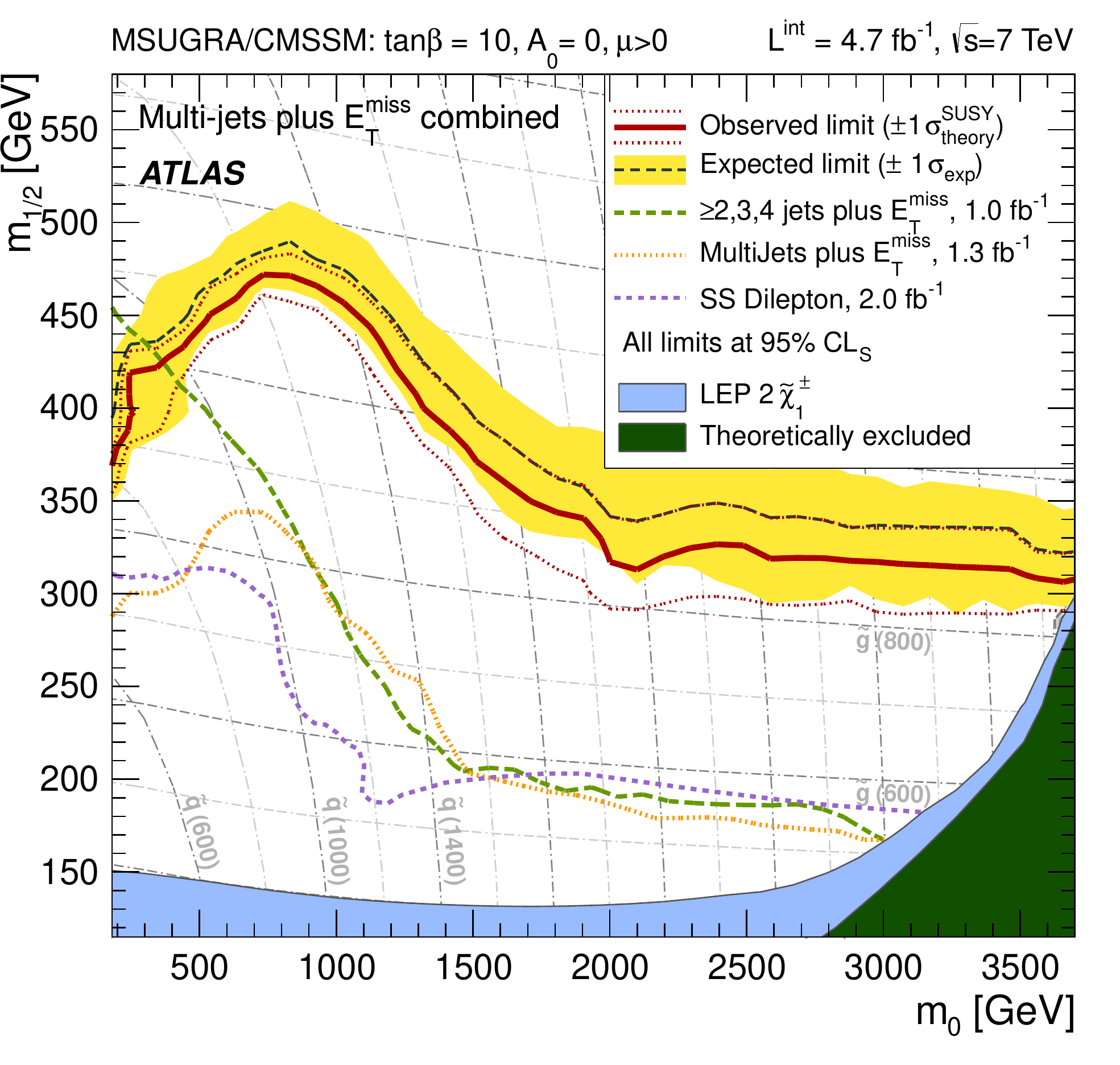}} \\
\subfigure[\label{fig:exclusion_combined:Gt} $\gluino - \ninoone$ simplified model]{\hspace{6mm}\includegraphics[width=0.69\linewidth]{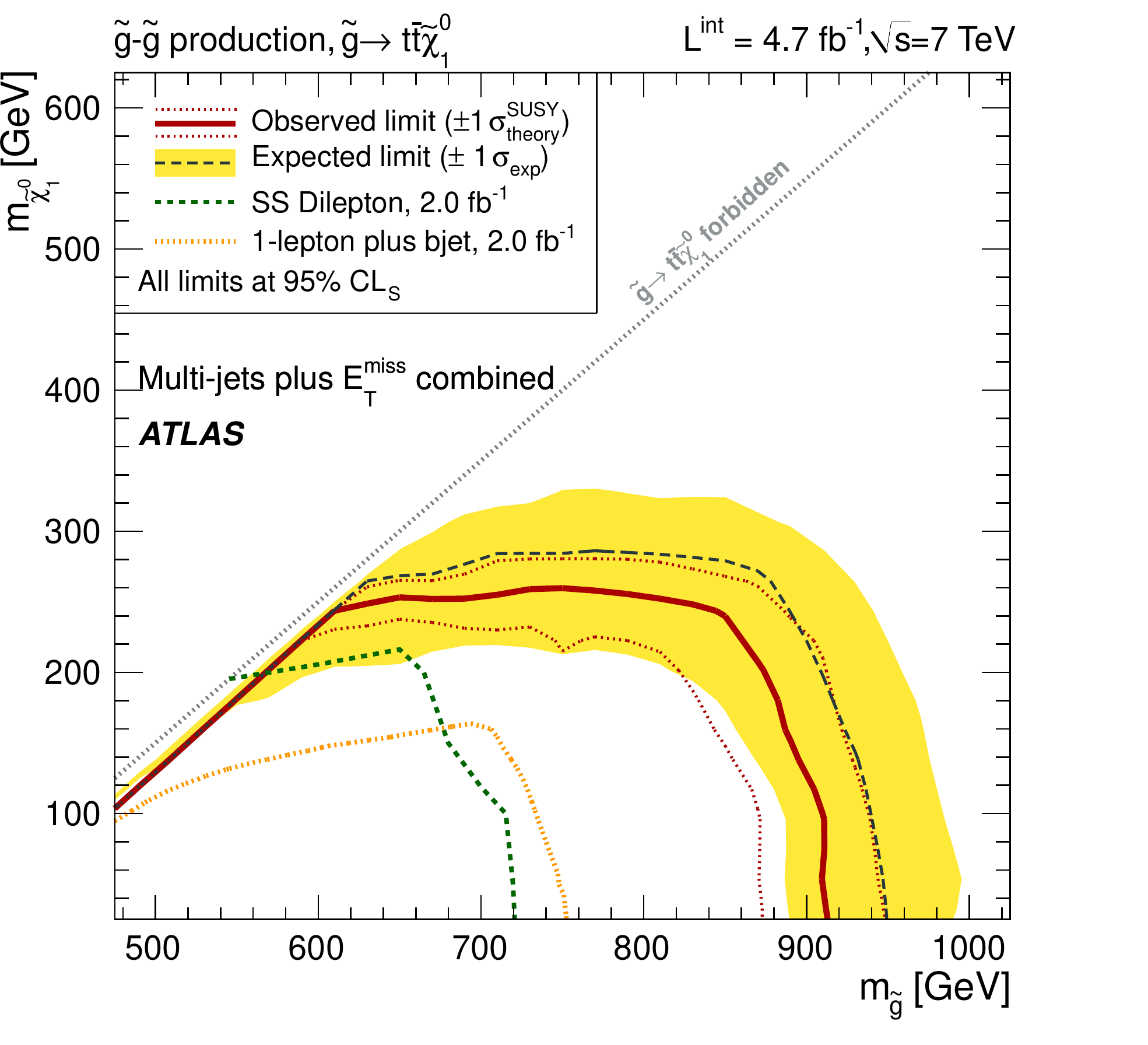}}
\caption{\label{fig:exclusion_combined}
  Combined 95\% CL exclusion curves for the 
  $\tan\beta=10$, $A_0=0$ and $\mu>0$ slice of \msugra{} \oursubref{fig:exclusion_combined:msugra} 
  and for the simplified gluino-neutralino model
  \oursubref{fig:exclusion_combined:Gt}. The
  dashed grey and solid red lines show the
  95\% CL expected and observed limits respectively,
  including all uncertainties except the theoretical signal
  cross section uncertainty (PDF and scale). The shaded yellow band around
  the expected limit shows its $\pm1\sigma$ range. The $\pm1\sigma$ lines around the observed limit represent the
  result produced when moving the signal cross section by $\pm1\sigma$
  (as defined by the PDF and scale uncertainties). 
  The contours on the \msugra{} model show values of the 
  mass of the gluino and the mean mass of the squarks in the first two generations.
  Exclusion limits are also shown from previous ATLAS searches
  with $\geq$2, 3 or 4 jets plus \met{}~\cite{Aad:2011ib}, multi-jets plus \met{}~\cite{Aad:2011qa} or with same-sign dileptons \cite{ATLAS-CONF-2012-004} and from LEP~\cite{lepsusy} in~\oursubref{fig:exclusion_combined:msugra}.
  The lower plot shows limits from ATLAS searches with same-sign dileptons~\cite{ATLAS-CONF-2012-004} or with one-lepton plus $b$-jet~\cite{ATLAS-CONF-2012-003}.
}
\end{figure*}

\figref{fig:metsig} shows
the \metsig{} distributions after applying the jet selections
for the six different signal regions (see \tabref{tab:sr})
prior to the final $\metsig{}> 4 \,\rootgev$ requirement.
\figref{fig:jetmult_signal} shows the jet multiplicity
distributions for the two different jet \pt{} thresholds
after the final \metsig{} requirement.
It should be noted that the signal regions are not exclusive. For example, in \figref{fig:metsig}, 
all plots contain the same event at $\metsig\sim11 \,\rootgev$.
The \leptonic{} backgrounds shown in the figures are those calculated from the
Monte Carlo simulation, using the MC calculation of the cross section and normalized to \ourintlumi{}.
The number of events observed in each of the six signal regions,
as well as their Standard Model background expectations are shown in \tabref{tab:results}.
Good agreement is observed between SM expectations and the data for all six signal regions.
Table~\ref{tab:results} also shows the 95\% confidence-level upper bound \nmaxbsm{} on the number of events originating from
sources other than the Standard Model, the corresponding upper limit $\sigmamaxbsm{} \times A \times \epsilon$ on the
cross section times efficiency within acceptance
(which equals the limit on the observed number of signal events divided by the luminosity)
and the $p$-value for the Standard-Model-only hypothesis ($p_{\rm SM}$).

In the absence of significant discrepancies, limits are set 
in the context of two supersymmetric (SUSY) models. 
The first is the $\tan\beta=10$, $A_0=0$ and $\mu>0$ slice of the \msugra{} parameter space. 
The second is a simplified SUSY model with only a gluino
octet and a neutralino $\ninoone$ within kinematic reach. 
Theoretical uncertainties on the SUSY signals are estimated as described 
in Section~\ref{sec:mc}.
Combined experimental
systematic uncertainties on the signal yield from jet energy scale, resolution,
and event cleaning are approximately $25\%$.
\hepdatablurb{}

The limit for each signal region is obtained by comparing the observed event count with
that expected from Standard Model background plus SUSY signal processes,
taking into account all uncertainties on the Standard Model expectation, including those which
are correlated between signal and background (for instance jet energy
scale uncertainties) and all but theoretical cross section uncertainties (PDF and scale) on the signal expectation.
The combined exclusion regions are obtained using the $CL_s$ prescription~\cite{clsread}, 
taking the signal region with the best expected limit
at each point in parameter space. 
The 95\% confidence level (CL) exclusion in the 
$\tan\beta=10$, $A_0=0$ and $\mu>0$ slice of 
\msugra{} is shown in \figref{fig:exclusion_combined}.
The $\pm$1\,$\sigma$ band surrounding the expected limit shows 
the variation anticipated from statistical fluctuations and systematic
uncertainties on SM and signal processes.
The uncertainties on the supersymmetric signal cross section from PDFs and higher-order terms
are calculated as described in \oursecref{sec:mc},
and the resulting signal cross section uncertainty is represented by $\pm 1\sigma$ lines on either side of the observed limit.\footnote{Previous analyses have a slightly different presentation of the effect of the signal cross section uncertainty. 
In Refs.~\cite{Aad:2011qa,Aad:2011ib,Aad:2011cw} the effect of the signal cross section uncertainty was folded into the displayed limits and so was not shown separately.}

The analysis substantially extends the previous exclusion 
limits~\cite{Aad:2011qa,Aad:2011ib,Aad:2011cw} for $m_0 > 500\GeV$.
For large $m_0$, the analysis becomes independent of the squark mass,
and the lower bound on the gluino mass is extended to almost $840\,\GeV$
for large $m_{\squark}$.\footnote{Limits on sparticle masses quoted in the text are those
from the lower edge of the $1\,\sigma$ signal cross section band rather than the central value of the observed limit,
so can be considered conservative.}
In the simplified model gluinos are pair-produced and
decay with unit probability to $t+\bar{t}+\ninoone$.
In this context, the 95\% CL exclusion bound on the gluino mass is
$870\,\GeV{}$ for neutralino masses up to $100 \GeV$.

\section{Summary}\label{sec:conclusions}

A search for new physics is presented using final states
containing large jet multiplicities in association with
missing transverse momentum.
The search uses the full 2011 $pp$ LHC data set taken at $\sqrt{s}=7\TeV$, collected with the ATLAS detector,
which corresponds to an integrated luminosity of $\ourintlumi$.

Six non-exclusive signal regions are defined.
The first three require at least seven, eight or nine jets, with $\pt{}>55\,\GeV$;
the latter three require at least six, seven or eight jets, with $\pt{}>80\,\GeV$.
In all cases the events are required to satisfy $\metsig>4 \rootgev$,
and to contain no isolated high-\pT{} electrons or muons.
Investigations on the enlarged data sample 
have resulted in improvements compared to 
a previous measurement using a similar strategy. 
In particular, inclusion of events with smaller
jet--jet separation increases the acceptance for signal models 
of interest by a factor two to five, without
without significantly increasing the systematic uncertainty.

The Standard Model multi-jet background is determined using 
a template-based method that exploits the invariance of \metsig{} under
changes in jet multiplicity,
cross-checked with a jet-smearing method that uses well reconstructed multi-jet seed events from data.
The other significant backgrounds ---
\ttbar{}~+~jets, $W$~+~jets and $Z$~+~jets --- are
determined using a combination of data-driven and Monte Carlo-based methods.

In each of the six signal regions, 
agreement is found between the Standard Model prediction and the data.
In the absence of significant discrepancies,
the results are interpreted as limits in the context of $R$-parity conserving supersymmetry.
Exclusion limits are shown for \mSUGRA{}, 
for which, for large $m_0$, gluino masses smaller than $840\GeV$
are excluded at the 95\% confidence level.
For a simplified supersymmetric model in 
which both of the pair-produced gluinos decay via the process
$\gluino \rightarrow t + \bar{t} + \ninoone$, gluino masses smaller than
about $870\GeV$ are similarly excluded for $\ninoone$ masses up to $100\,\GeV$.

\FloatBarrier

\section{Acknowledgments}




We thank CERN for the very successful operation of the LHC, as well as the
support staff from our institutions without whom ATLAS could not be
operated efficiently.

We acknowledge the support of ANPCyT, Argentina; YerPhI, Armenia; ARC,
Australia; BMWF, Austria; ANAS, Azerbaijan; SSTC, Belarus; CNPq and FAPESP,
Brazil; NSERC, NRC and CFI, Canada; CERN; CONICYT, Chile; CAS, MOST and NSFC,
China; COLCIENCIAS, Colombia; MSMT CR, MPO CR and VSC CR, Czech Republic;
DNRF, DNSRC and Lundbeck Foundation, Denmark; EPLANET and ERC, European Union;
IN2P3-CNRS, CEA-DSM/IRFU, France; GNAS, Georgia; BMBF, DFG, HGF, MPG and AvH
Foundation, Germany; GSRT, Greece; ISF, MINERVA, GIF, DIP and Benoziyo Center,
Israel; INFN, Italy; MEXT and JSPS, Japan; CNRST, Morocco; FOM and NWO,
Netherlands; RCN, Norway; MNiSW, Poland; GRICES and FCT, Portugal; MERYS
(MECTS), Romania; MES of Russia and ROSATOM, Russian Federation; JINR; MSTD,
Serbia; MSSR, Slovakia; ARRS and MVZT, Slovenia; DST/NRF, South Africa;
MICINN, Spain; SRC and Wallenberg Foundation, Sweden; SER, SNSF and Cantons of
Bern and Geneva, Switzerland; NSC, Taiwan; TAEK, Turkey; STFC, the Royal
Society and Leverhulme Trust, United Kingdom; DOE and NSF, United States of
America.

The crucial computing support from all WLCG partners is acknowledged
gratefully, in particular from CERN and the ATLAS Tier-1 facilities at
TRIUMF (Canada), NDGF (Denmark, Norway, Sweden), CC-IN2P3 (France),
KIT/GridKA (Germany), INFN-CNAF (Italy), NL-T1 (Netherlands), PIC (Spain),
ASGC (Taiwan), RAL (UK) and BNL (USA) and in the Tier-2 facilities
worldwide.


\bibliographystyle{JHEP}
\bibliography{MultiJets2012}

\appendix

\section{Event displays}\label{sec:app}

A display of an event that passes the 9j55 and 7j80 signal region selections can be found in 
\figref{fig:evd}. 
A display of an event that passes all signal region selections can be found in 
\figref{fig:evd1}. 

\begin{figure*}
\centering
\subfigure[]{\includegraphics[width=0.95\linewidth]{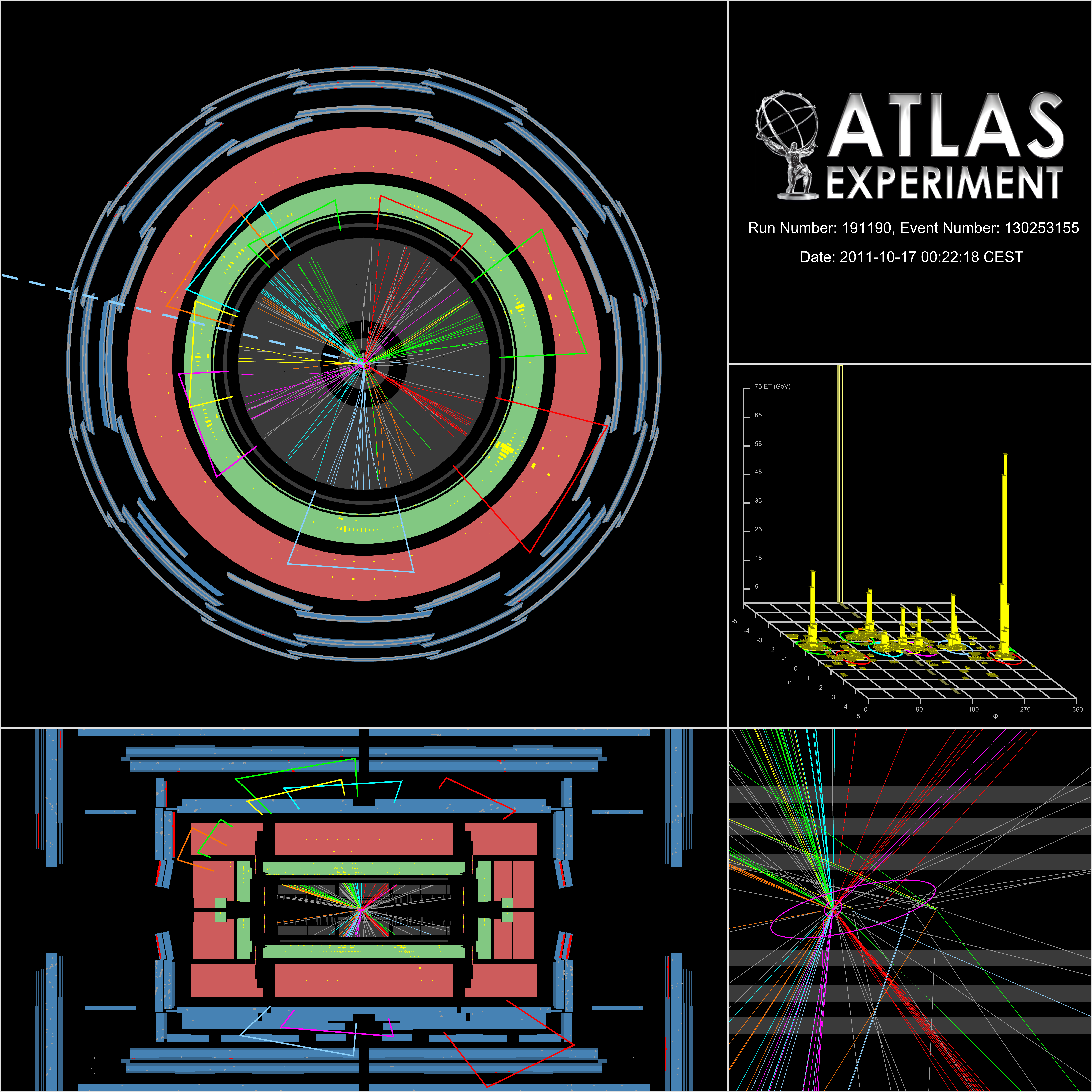} }
\caption{\label{fig:evd}
  A display of an event 
  which passes the 9j55 and 7j80 signal region selections. The event
  has $\metsig{}$ of $4.1\,\rootgev{}$, $\rm{H_T}$ of 1.47~TeV and 
  \met{} of 157\,\GeV{}.
}
\end{figure*}

\begin{figure*}
\centering
\subfigure[]{\includegraphics[width=0.95\linewidth]{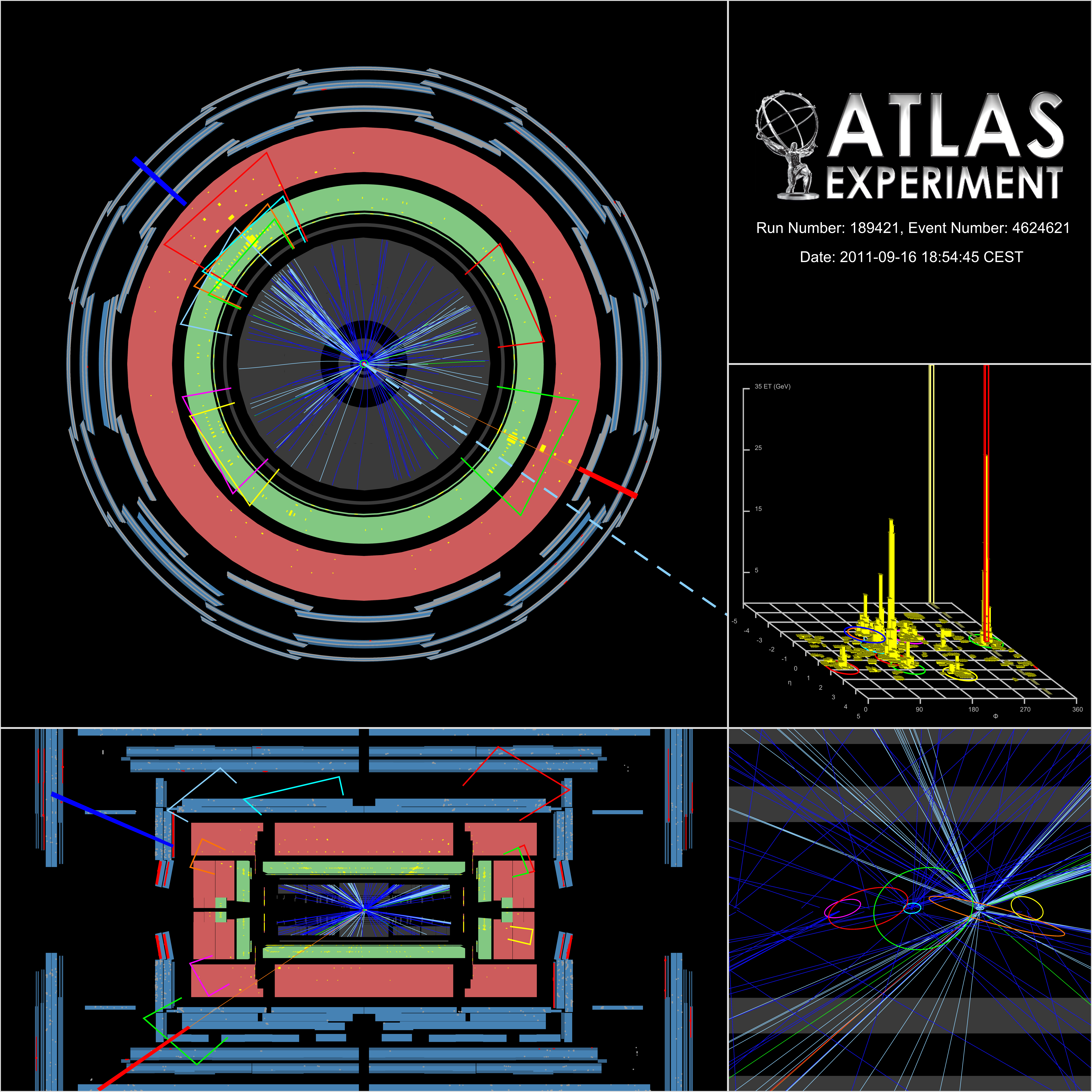}}
\caption{\label{fig:evd1}
  A display of an event
  which passes all signal region selections. The event has $\metsig{}$ of $11.6\,\rootgev{}$, $\rm{H_T}$ of 1.17~TeV and
  \met{} of 397\,\GeV{}. One of the jets, with $\rm{p_T}$ of 107\,\GeV{} is $b$ tagged. The event also
  contains a muon with $\rm{p_T}$ of 90\,\GeV{}, overlapping with a
  jet. 
}
\end{figure*}

\clearpage
\begin{flushleft}
{\Large The ATLAS Collaboration}

\bigskip

G.~Aad$^{\rm 48}$,
B.~Abbott$^{\rm 111}$,
J.~Abdallah$^{\rm 11}$,
S.~Abdel~Khalek$^{\rm 115}$,
A.A.~Abdelalim$^{\rm 49}$,
O.~Abdinov$^{\rm 10}$,
B.~Abi$^{\rm 112}$,
M.~Abolins$^{\rm 88}$,
O.S.~AbouZeid$^{\rm 158}$,
H.~Abramowicz$^{\rm 153}$
% ...
\bigskip

% ...
$^{10}$ Institute of Physics, Azerbaijan Academy of Sciences, Baku, Azerbaijan\\
$^{11}$ Institut de F\'isica d'Altes Energies and Departament de F\'isica de la Universitat Aut\`onoma  de Barcelona and ICREA, Barcelona, Spain\\
% ...
$^{48}$ Fakult\"{a}t f\"{u}r Mathematik und Physik, Albert-Ludwigs-Universit\"{a}t, Freiburg i.Br., Germany\\
$^{49}$ Section de Physique, Universit\'e de Gen\`eve, Geneva, Switzerland\\
% ...
$^{88}$ Department of Physics and Astronomy, Michigan State University, East Lansing MI, United States of America\\
% ...
$^{111}$ Homer L. Dodge Department of Physics and Astronomy, University of Oklahoma, Norman OK, United States of America\\
$^{112}$ Department of Physics, Oklahoma State University, Stillwater OK, United States of America\\
% ...
$^{115}$ LAL, Universit\'e Paris-Sud and CNRS/IN2P3, Orsay, France\\
% ...
$^{153}$ Raymond and Beverly Sackler School of Physics and Astronomy, Tel Aviv University, Tel Aviv, Israel\\
% ...
$^{158}$ Department of Physics, University of Toronto, Toronto ON, Canada\\
% ...
$^{p}$ Also at Group of Particle Physics, University of Montreal, Montreal QC, Canada\\
% ...
$^{*}$ Deceased\end{flushleft}


\end{document}